\documentclass[a4paper,11pt]{article}
\pdfoutput=1 % if your are submitting a pdflatex (i.e. if you have
\usepackage{jinstpub} % for details on the use of the package, please
\usepackage{hepparticles}
\usepackage{heppennames}
\usepackage{threeparttable}
\usepackage{subcaption}
\usepackage{tikz}
\usepackage{hyperref}
\usepackage{siunitx}
\graphicspath{{figures/}}
\title{\boldmath Characterization of Silicon-Photomultipliers for a Cosmic Muon Veto detector}

\author[a,b,1]{Mamta Jangra,\note{Corresponding author.}}
\author[b]{Gobinda Majumder,}
\author[b]{Mandar Saraf,}
\author[b]{B. Satyanarayana,}
\author[b]{R.R. Shinde,}
\author[b]{Suresh S Upadhya,}
\author[c]{Vivek M Datar,}
\author[d]{Douglas A. Glenzinski,}  
\author[d]{Alan Bross,}
\author[d]{Anna Pla-Dalmau,}
\author[e]{Vishnu V. Zutshi,}
\author[f]{Robert Craig Group} 
\author[f]{and E Craig Dukes }

\affiliation[a]{Homi Bhabha National Institute, Mumbai-400094, India}
\affiliation[b]{Tata Institute of Fundamental Research, Mumbai-400005, India}
\affiliation[c]{The Institute of Mathematical Sciences, Chennai-600113, India}
\affiliation[d]{Fermi National Accelerator Laboratory, IL 60510, United States}
\affiliation[e]{Northern Illinois University, IL 60510, United States}
\affiliation[f]{Virginia University, VA, United States}
\emailAdd{mamta.jangra@tifr.res.in}

\abstract{A Cosmic Muon Veto (CMV) detector using extruded scintillators is being designed around the mini-Iron Calorimeter detector at the transit campus of the India-based Neutrino Observatory at Madurai for measuring its efficiency at shallow depth underground experiments. The scintillation signal is transmitted through a Wavelength Shifting (WLS) fibre and readout by Hamamatsu Silicon-Photomultipliers (SiPMs). A Light Emitting Diode (LED) system is included on the front-end readout for in-situ calibration of the gain of each SiPM. A characterization system was developed for the measurement of gain and choice of the overvoltage ($V_{ov}$) of SiPMs using LED as well as a cosmic muon telescope. The $V_{ov}$ is obtained by studying the noise rate, the gain of the SiPM, and the muon detection efficiency. In case of any malfunction of the LED system during the operation, the SiPM can also be calibrated with the noise data as well as using radioactive sources. This paper describes the basic characteristics of the SiPM and the comparison of the calibration results using all three methods, as well as the $V_{ov}$ of the SiPMs and muon selection criteria for the veto detector.}

\begin{document}
\maketitle
\flushbottom
\raggedbottom
\section{Introduction}
\label{intro}
A 51\,kton magnetized Iron Calorimeter (ICAL) was proposed~\cite{inoreport} at the underground laboratory of India-based Neutrino Observatory (INO) to precisely measure the parameters of atmospheric neutrinos, mainly to study the effect of matter on neutrino oscillations~\cite{inowhitepaper}. The underground laboratory (with rock cover of more than 1\,km in all directions) along with ICAL is planned to be located in Bodi West hills at Theni (\ang{9;57;50.1}\,N, \ang{77;16;21.8}\,E), India. ICAL will consist of three modules, where each of the modules will contain 150 layers of Resistive Plate Chambers (RPCs) interfaced between 5.6\,cm thick iron plates. Due to the low interaction cross-section of neutrino, there are only a handful of neutrino events in a day whereas there are a large number of muons $\sim$ $3 \times 10^{8}$ /day coming from a cosmic shower at the surface for the proposed ICAL detector~\cite{muonfluxcalculation}.
These cosmic muons act as a huge background for the detection of muons arising from neutrino interactions. Placing a neutrino detector under a rock cover of 1km in all directions reduces the cosmic muon flux by a factor of $\sim$ $10^{6}$. At a depth of $\sim$ 100\,m or so, muons will be suppressed by a factor of $10^{2}$. To achieve a rejection factor $10^{6}$, an active cosmic muon veto system with an efficiency of atleast 99.99$\%$ must be built around such a shallow depth detector.\\
The prototype detector of the ICAL i.e., mini-ICAL is currently in operation at IICHEP, Madurai~\cite{gmsir}. The mini-ICAL consists of twenty layers of $2\,m \times 2\,m$ glass RPCs sandwiched between 11 layers of 5.6\,cm thick iron plates. The detector is magnetised to a field of about 1.4\,T using the iron plates and two copper coils of 18 turns each. To suppress the cosmic muon background, it is planned is to cover the top and three sides of the mini-ICAL with an active veto detector for the detection of the cosmic ray muons and estimate the cosmic muon veto efficiency of the detector, using the same concept as of mu2e collaboration~\cite{mu2etdr}. The goal is to build an active veto detector with an efficiency of 99.99$\%$ and a fake rate of less than $10^{-5}$. 

\begin{figure} [h!]  
\centering
\includegraphics[height=9.5cm,width=14.0cm]{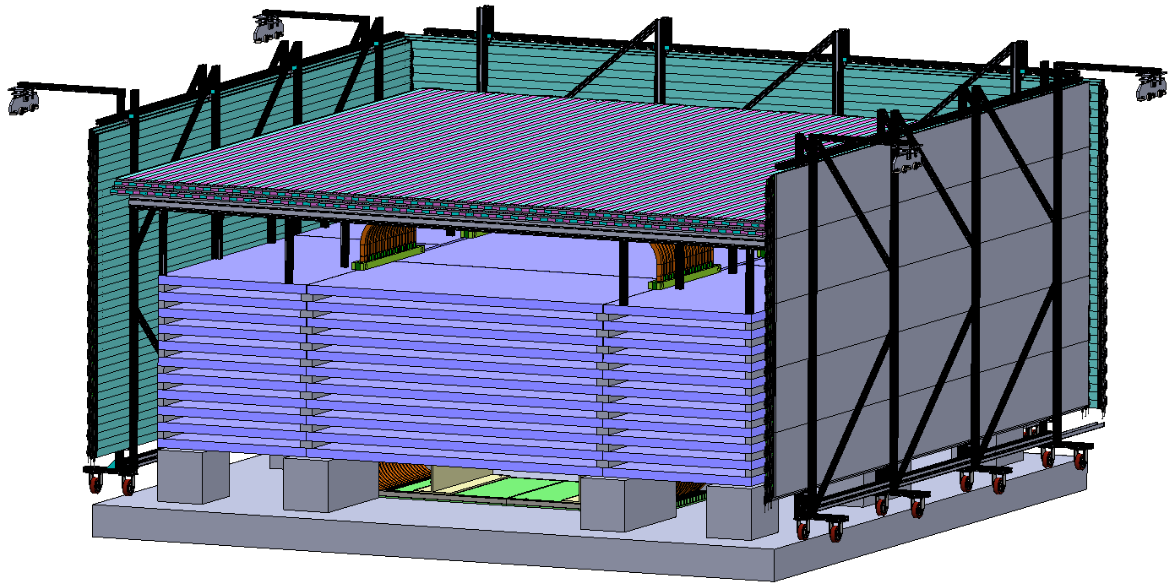}
\caption{Overall layout of the CMV detector around mini-ICAL.}
\label{fig:cmvdetector}
\end{figure}

The CMV detector for mini-ICAL will have four layers of extruded scintillators~\cite{lowcost} of size ($460\,cm \times 5\,cm \times 2\,cm$) on top and three layers of ($460\,cm \times 5\,cm \times 1\,cm$) on three sides of the mini-ICAL as shown in Fig.~\ref{fig:cmvdetector}. The front side is not covered with veto layers due to the maintenance and troubleshooting of the mini-ICAL. In all, the veto system will comprise of more than 700 extruded plastic scintillators. The surface area of each layer on top is ($460\,cm \times 440\,cm$) and is ($440\,cm \times 200\,cm$) for each of the side layers. The scintillation signal will be collected through WLS fibres embedded in the extruded scintillator and the light from both ends of the fibres will be readout using $2\,mm \times 2\,mm$ Hamamatsu SiPMs (S13360-2050VE). \\
This paper describes the characterization of the extruded scintillators and SiPMs which are the main components of cosmic muon veto setup. These studies were performed as a part of the R$\&$D program for the cosmic muon veto system around mini-ICAL. A veto efficiency of 99.98$\%$ was measured with a small detector which was setup using plastic scintillators with readout by PMTs~\cite{nehapanchal1}.

\section{Experimental setup}
\label{exptsetup}
\begin{figure}[htbp]
\centering
\begin{minipage}{18pc}
\includegraphics[height=5cm,width=7.5cm]{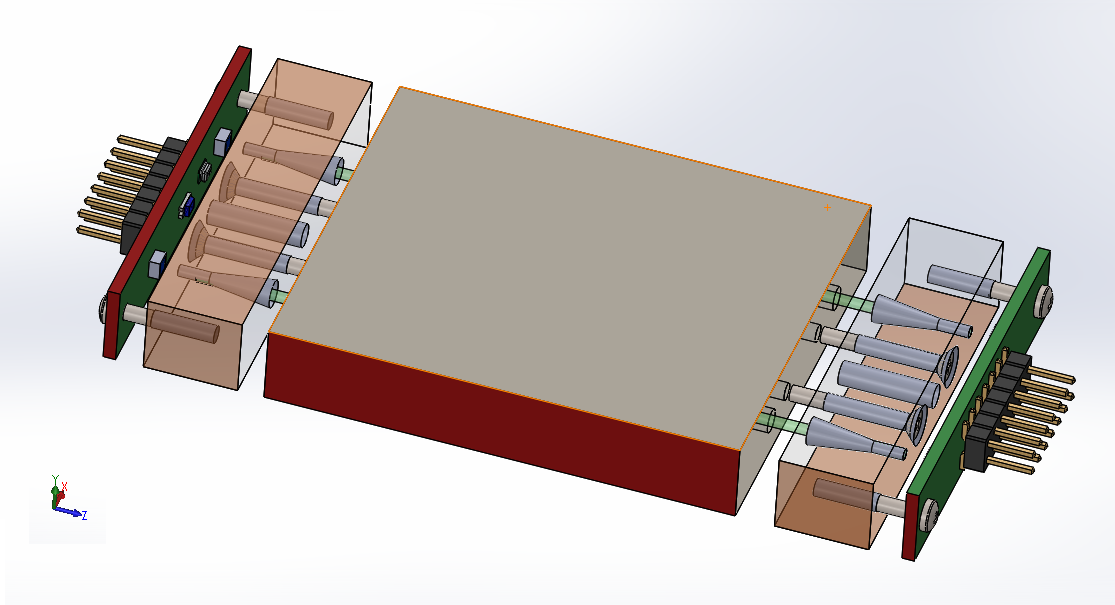}
\caption{\label{fig:CMB1}Exploded view of extruded scintillator, the fibre guide bar (FGB) assembly and SiPM mounting board (SMB).}
\end{minipage}
\hspace{0.75cm}
\begin{minipage}{15pc}
\vspace*{-0.5cm}
\begin{tikzpicture}
\node[above right] (img) at (0,0) {\includegraphics[height=5cm,width=6.cm]{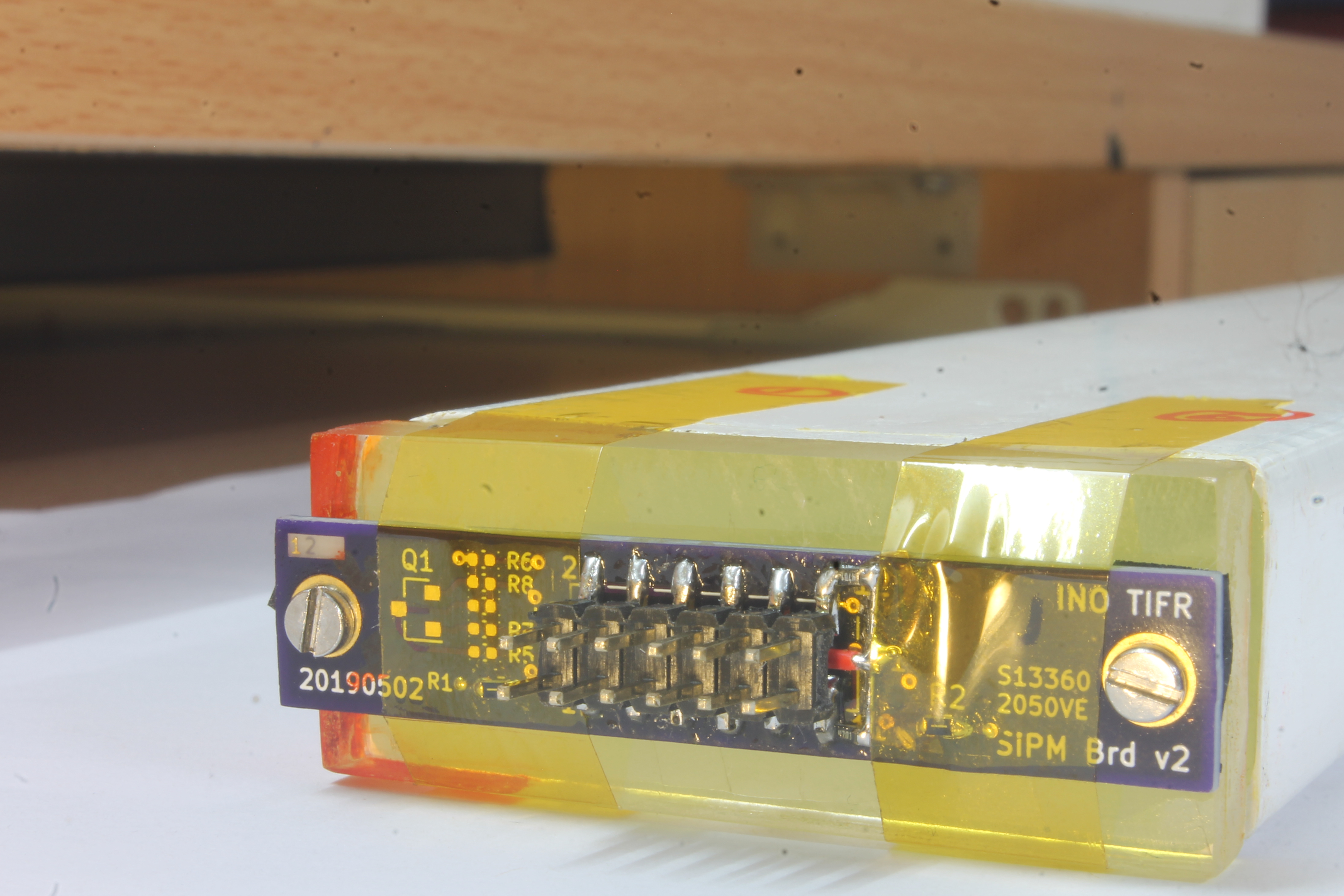}};
\color{red}
%\node at (25pt,15pt) {$Pins$};
%\node at (60pt,70pt) {$\searrow$};
\node at (68pt,28pt) {$\uparrow$};
\node at (75pt,28pt) {$\uparrow$};
\node at (82pt,28pt) {$\uparrow$};
\node at (68pt,50pt) {$\downarrow$};
\node at (75pt,50pt) {$\downarrow$};
\node at (82pt,50pt) {$\downarrow$};
\node at (68pt,15pt) {$1$};
\node at (68pt,63pt) {$2$};
\node at (75pt,15pt) {$3$};
\node at (75pt,63pt) {$4$};
\node at (82pt,15pt) {$5$};
\node at (82pt,63pt) {$6$};
\end{tikzpicture}
\caption{\label{fig:CMB2}SiPM board mounted on a 2\,cm thick extruded scintillator.}
\end{minipage}
\end{figure}

\begin{figure} [htbp]  
\centering
\includegraphics[height=7.5cm,width=12.25cm]{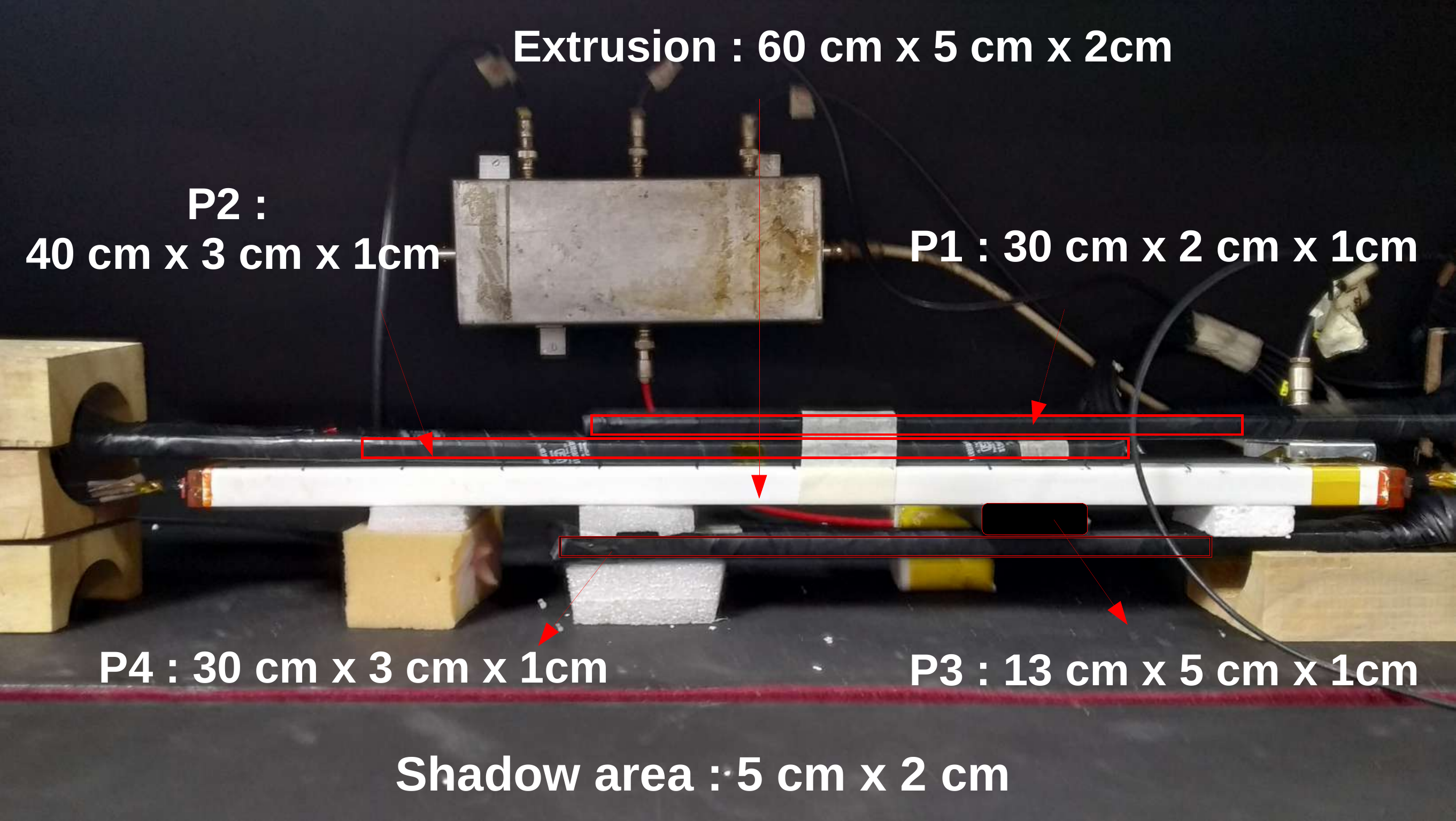}
\caption{Experimental setup for studies with cosmic ray muons.}
\label{fig:blackbox}
\end{figure}

Each extruded scintillator contains two elliptical holes
% in Fig.~\ref{fig:cmvdetector} showing cross sectional view 
throughout the length into which WLS fibres of diameter 1.4\,mm were inserted~\cite{Mu2epaper}. A fibre guide bar (FGB) made of acrylic is glued on both sides of the extruded scintillator as shown in Fig.~\ref{fig:CMB1}. Each SiPM mounting board (SMB) as shown in the Fig.~\ref{fig:CMB2} contains two SiPMs. The SMB provides the required mechanical support and secures electrical connection to the SiPMs and is mated to the FGB in such a way that the SiPM surface will be in close proximity with the WLS fibre end. Since each counter contains two WLS fibres, there will be a total of four SiPMs to readout light on both sides of the scintillator.\\
The complete experimental setup as shown in the Fig.~\ref{fig:blackbox} is placed in a cuboidal black box to create a light tight environment for the setup. The studies were performed on extruded scintillators of two different dimensions ($60\,cm \times 5\,cm \times 2\,cm$) and ($60\,cm \times 5\,cm \times 1\,cm$). The SiPM on SMB is powered using a Keithley 2400 Sourcemeter. Each SiPM has a common supply voltage given through pin numbers (1, 2). A total current of 0.2\,$\mu$A is drawn at 54\,V at room temperature. The output signal is connected to a digital storage oscilloscope via coaxial cables from pin numbers (3, 4) and (5, 6) on the SMB. The signals are acquired at sampling rates of (1\,GSa/s to 20\,GSa/s).

\subsection{Trigger}
The cosmic muon trigger signal is generated by the four-fold time coincidence of plastic scintillator paddles with photomultiplier tube readout. The coincidence was made using a NIM quad logic unit (Lecroy 365AL). The paddles, P1 ($30\,cm \times 2\,cm \times 1\,cm$), P2 ($30\,cm \times 3\,cm \times 1\,cm$), P3 ($13\,cm \times 5\,cm \times 1\,cm$) and P4 ($40\,cm \times 3\,cm \times 1\,cm$ ) are used for generating trigger with an effective telescope window of $5\,cm \times 2\,cm$, which is shown in Fig.~\ref{fig:blackbox}. The extruded scintillator with SiPM readout is placed in the middle of the trigger paddles such that the telescope window is aligned with the centre of the extruded scintillator under test. A time coincidence of shaped (40\,ns) logic signals from four scintillator paddles is used to indicate the passage of a cosmic ray muon through the detector setup. To increase the trigger rate, particularly, for the measurement of cosmic muon efficiency, P3 is replaced by a large area scintillator so that the telescope window increases to $\sim$ ($25\,cm \times 2\,cm$).\\
As stated above, besides cosmic ray muons, other sources or ways are used for calibration studies of the SiPM. The following triggers are setup for the data collection during these studies.
\begin{itemize}
\item For the LED calibration studies, the LED driver is used to generate the trigger.
\item For the collection of noise data, a randomly generated trigger is used. % For calibration of SiPM using noise data, one SiPM is used with a very low threshold to trigger the system, and data from the remaining three are studied.
\item For calibration of SiPM using a radioactive source, one SiPM is used to trigger the system with a threshold of a single photoelectron (pe) and data from the remaining three SiPM's are recorded.
\item For the calculation of correlated noise rates, a trigger on consecutive signal peaks with a threshold of half a photoelectron is used with an amplifier gain 25.
\end{itemize}
For the LED calibration and noise data, the SiPMs are disconnected from the extruded scintillator and WLS fibres.

\section{Calibration of SiPMs using LED}
\label{ledcalib}
The standalone SiPMs were calibrated using an LED system. An SP5601 CAEN LED system~\cite{CAEN} is used for SiPM testing and characterization. The LED system consists of an ultrafast LED driver with tunable frequency and intensity, and a pulse width of a few $ns$. It can provide a light burst with a tunable intensity varying from a few to more than tens of photons required for calibration purposes. A cubical wooden black box of side $\sim$10\,cm is used for SiPM testing with LED source. An arrangement is made inside the box with a slider to support the SMB holding the SiPM and also to provide the required electrical connections. The optical fibre from the LED driver is terminated on one of the faces of the black box such that the SiPM will directly face the LED flash. The LED pulse is synchronized with a trigger by an external pulser. The SiPM output is taken from the SMB to the oscilloscope using a coaxial cable.  \\
Each SiPM under test has a total of 1584 pixels. For analog SiPMs (common readout for all pixels), the overall signal of the SiPM will be proportional to the sum of the individual pixel's signal which are fired~\cite{sipmpaper2}. With each trigger, a handful of photons which are emitted from the LED system and generate electron-hole pairs in the SiPM's microcells. An avalanche is produced in every corresponding microcell due to the applied bias voltage ($V_{bias}$) and the resultant charge is measured. The integrated charge is calculated using the equation:

\begin{figure} [htbp]  
\centering
\includegraphics[height=8cm]{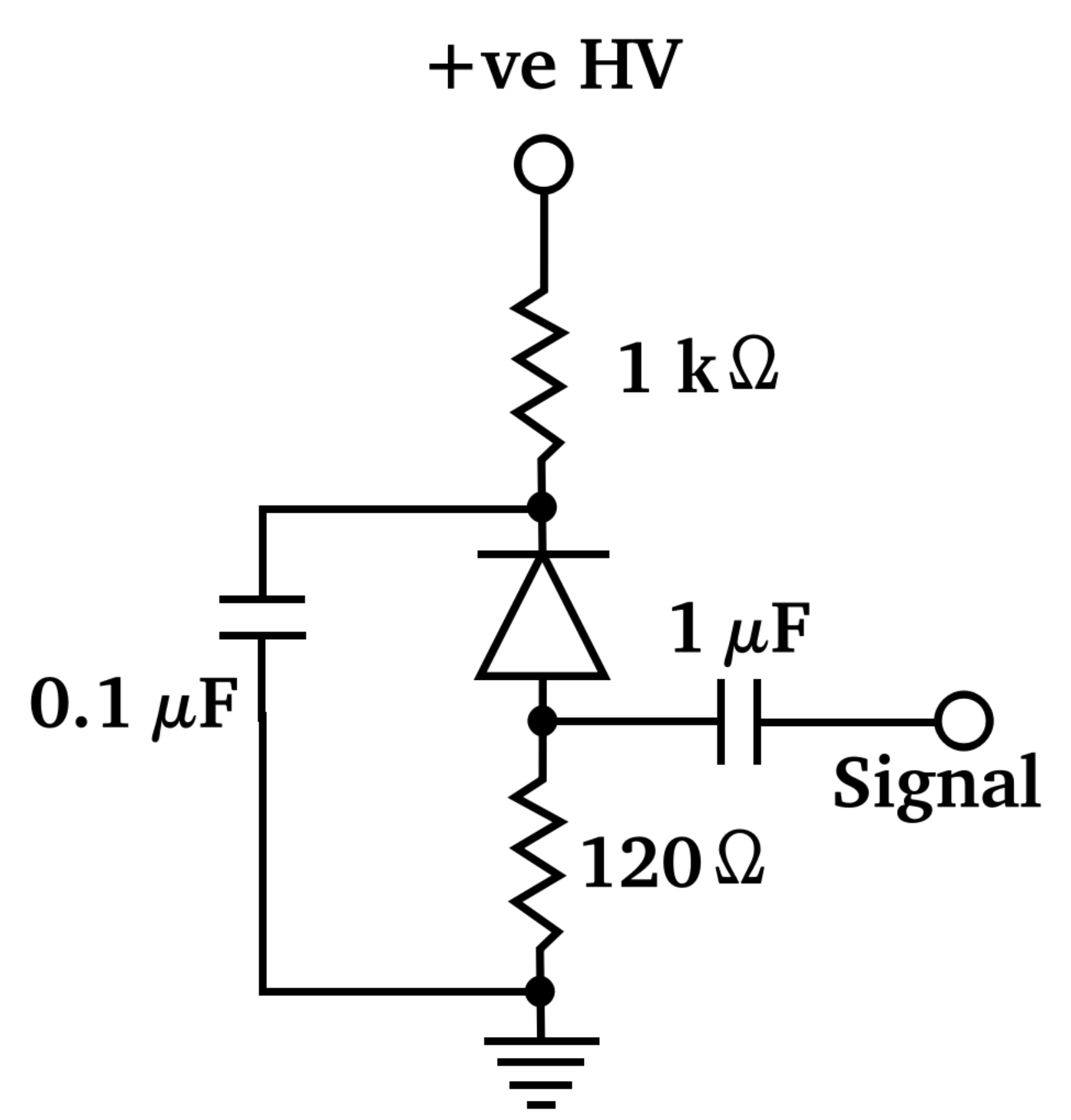}
\caption{Operating circuit diagram of SiPM.}
\label{fig:sipmckt}
\end{figure}

\begin{equation}
\label{eq:chargeintg}
q =  \frac{1}{R} \int_{t_{0}}^{t_{1}} V(t) dt
\end{equation}

where R = 120\,$\Omega$ as illustrated in Fig.~\ref{fig:sipmckt}. A decoupling capacitor (0.1\,$\mu$F) helps to supply a stable bias voltage by filtering out any noise that may come from the power supply. The series resistance (1\,K$\Omega$) is chosen to limit large current as well as to have a faster recovery time. To read out the SiPM from the standard output, the photocurrent generated on detection of photons needs to be converted to a voltage. This can be achieved using a series resistor (120\,$\Omega$), which also sets the transimpedance gain. The SiPM is a.c. coupled using 1\,$\mu$F capacitor to the electronics.

\begin{figure} [htbp]  
\centering
\includegraphics[height=6.75cm,width=15.cm]{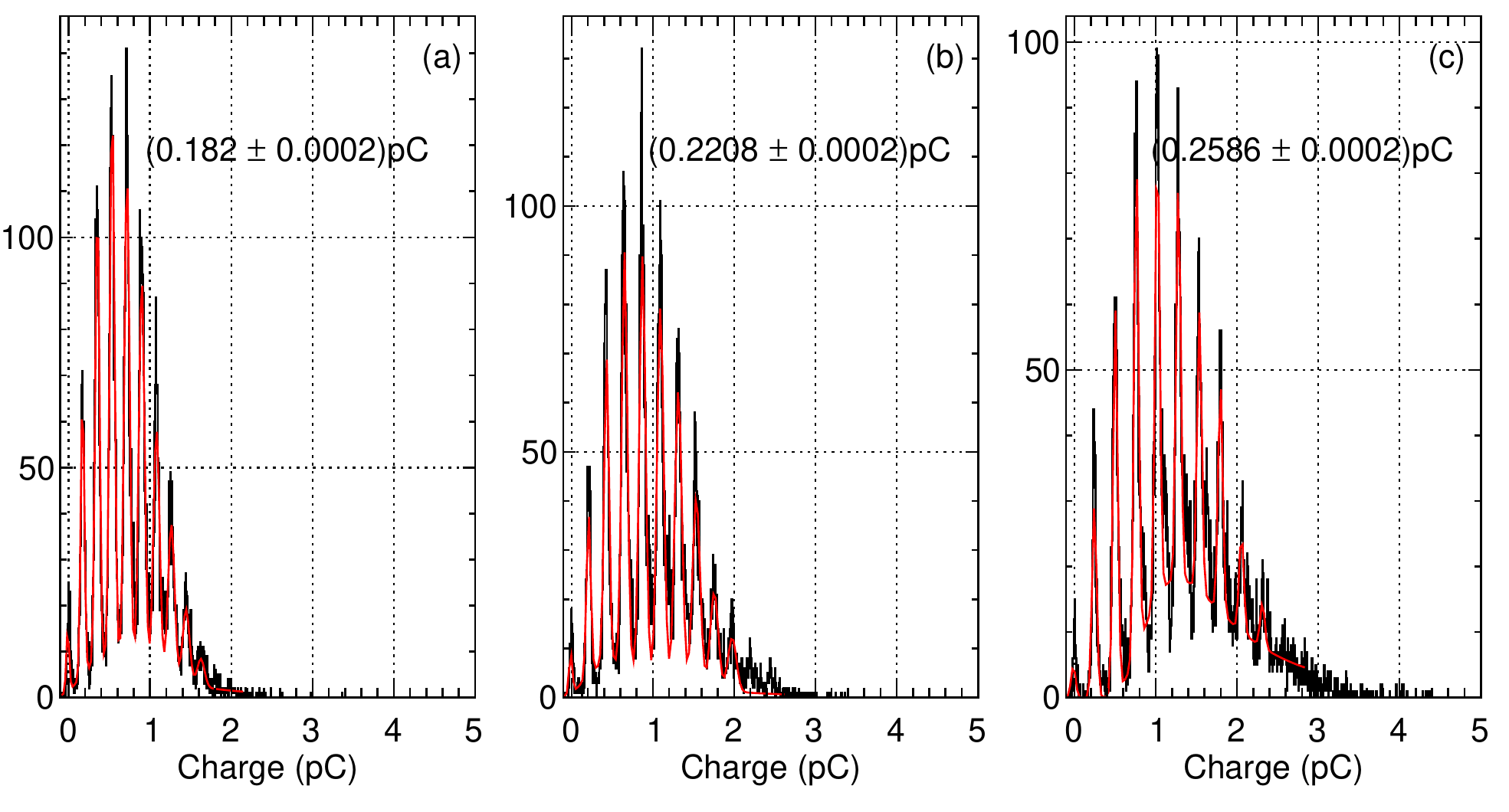}
\caption{Photoelectron peaks and average gap between consecutive photoelectron peaks at $V_{bias}$ = (a) 53.5\,V (b) 54\,V and (c) 54.5\,V.}
\label{fig:sipmpeakled}
\end{figure}

The SiPM response was measured at different values of $V_{ov}$ as shown in Fig.~\ref{fig:sipmpeakled}. Clear and distinct peaks corresponding to different photoelectrons can be seen. The integrated charge within 100\,ns is fitted with a function,

\begin{equation}
\label{eq:landgaus}
f(q) = Landau(q) + \sum_{n=0}^{N} p_{n} \times exp\Bigg[-\frac{(q-n\,\mu)^{2}}{2\sigma^2}\Bigg]
\end{equation}

where N is the number of photoelectron peaks, $p_{n}$ is the peak height of the $n^{th}$ photoelectron, $\mu$ is the gain of single photoelectron of the SiPM and $\sigma$ is the width of the peaks. There is no physics behind the Landau function for the continuous background distribution. The observed shape matched with the Landau function, thus it is used here. For an ideal case, there should be a sharp peak at different photoelectrons. But due to intrinsic electronic noise as well as non-uniformity of the SiPM, both within the pixel and pixel to pixel, the signal is broadened. The number of peaks depends on the intensity of the LED and the gap between two peaks depends on the gain of SiPM for a particular $V_{ov}$ at the given temperature. Due to self-absorption and crosstalk within the SiPM device, the height of the peak does not follow the Poisson distribution. There is a noticeable small bump on the right side of each photoelectron peak, which is mainly due to the afterpulse (multiple signals within the same pixel). At very low $V_{ov}$, events outside the peak are mainly due to noise in the electronic circuits, which is not negligible in comparison to the gain of a single photoelectron whereas at higher $V_{ov}$, as stated above, this is mainly due to afterpulses.

\begin{figure} [htbp]  
\centering
\includegraphics[height=6.75cm,width=11cm]{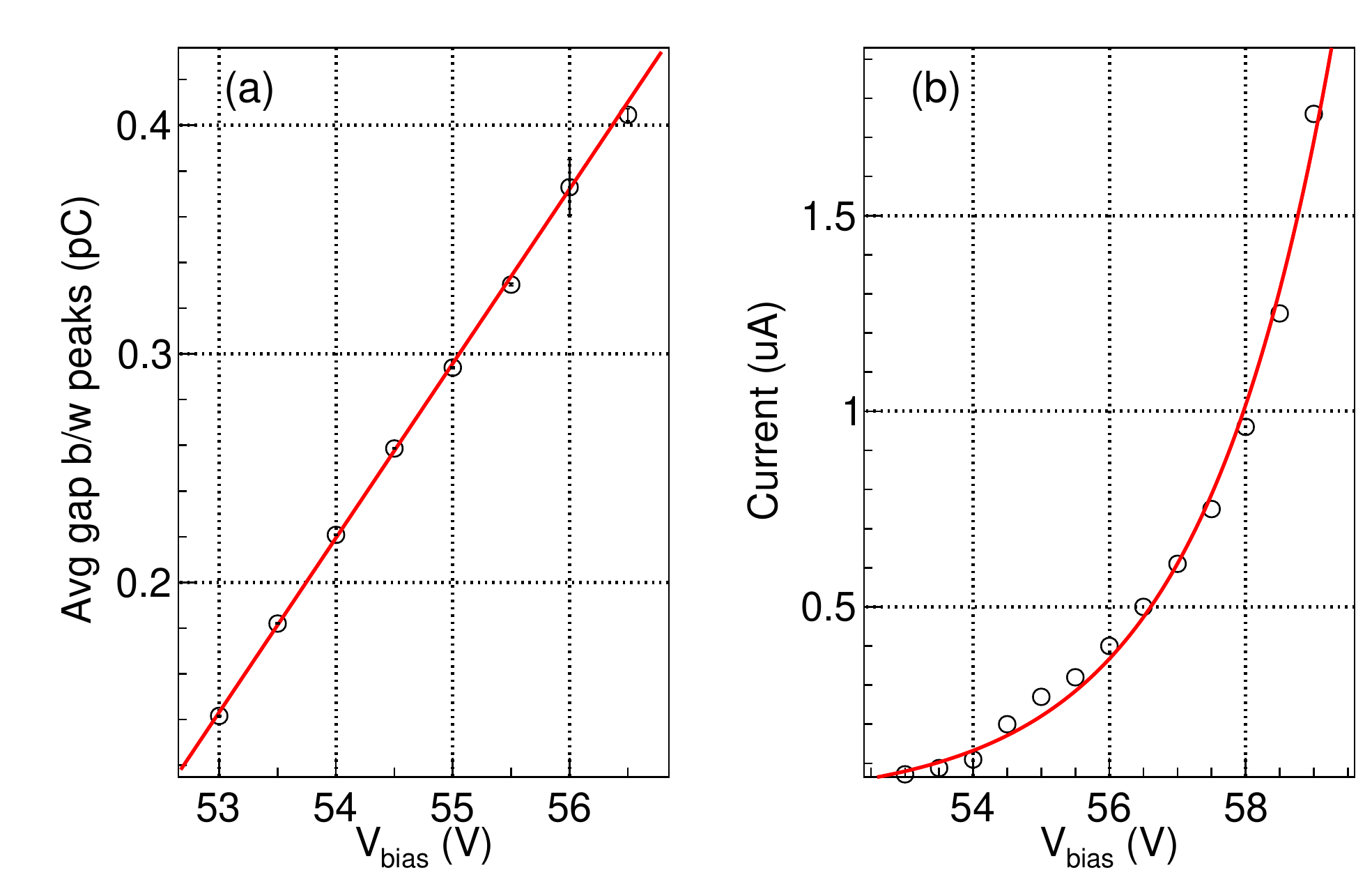}
\caption{(a) $V_{bias}$ versus average gap between two consecutive photoelectron peaks and (b) $V_{bias}$ versus dark current through an individual SiPM.}
\label{fig:calibration}
\end{figure}

The breakdown voltage ($V_{bd}$) is estimated from Fig.~\ref{fig:calibration}\color{blue}a\color{black}\hspace{0.1cm} and is found to be nearly (51.114 $\pm$ 0.005)\,V for this SiPM and most of the SiPMs under test show similar values within $\pm$0.5\,V\,\footnote{For simplicity, we assume the breakdown voltage, $V_{bd}$=51\,V and in the following we will use the convention of voltage as overvoltage ($V_{ov}$), which is the difference of the applied bias voltage $V_{bias}$ and $V_{bd}$.}. As expected, the dark current increases exponentially with $V_{bias}$ as shown in Fig.~\ref{fig:calibration}\color{blue}b\color{black}. There is no extra control of temperature and humidity other than the air conditioning system in the room, which is within (25 $\pm$ 0.5)$^{\circ}$ and (35 $\pm$ 5)\% respectively.

\section{Dependence of noise rate on overvoltage on SiPM}
\label{noiserate}
\begin{figure}[htbp]
\centering
\begin{minipage}{16pc}

\includegraphics[height=5.3cm, width=5.75cm]{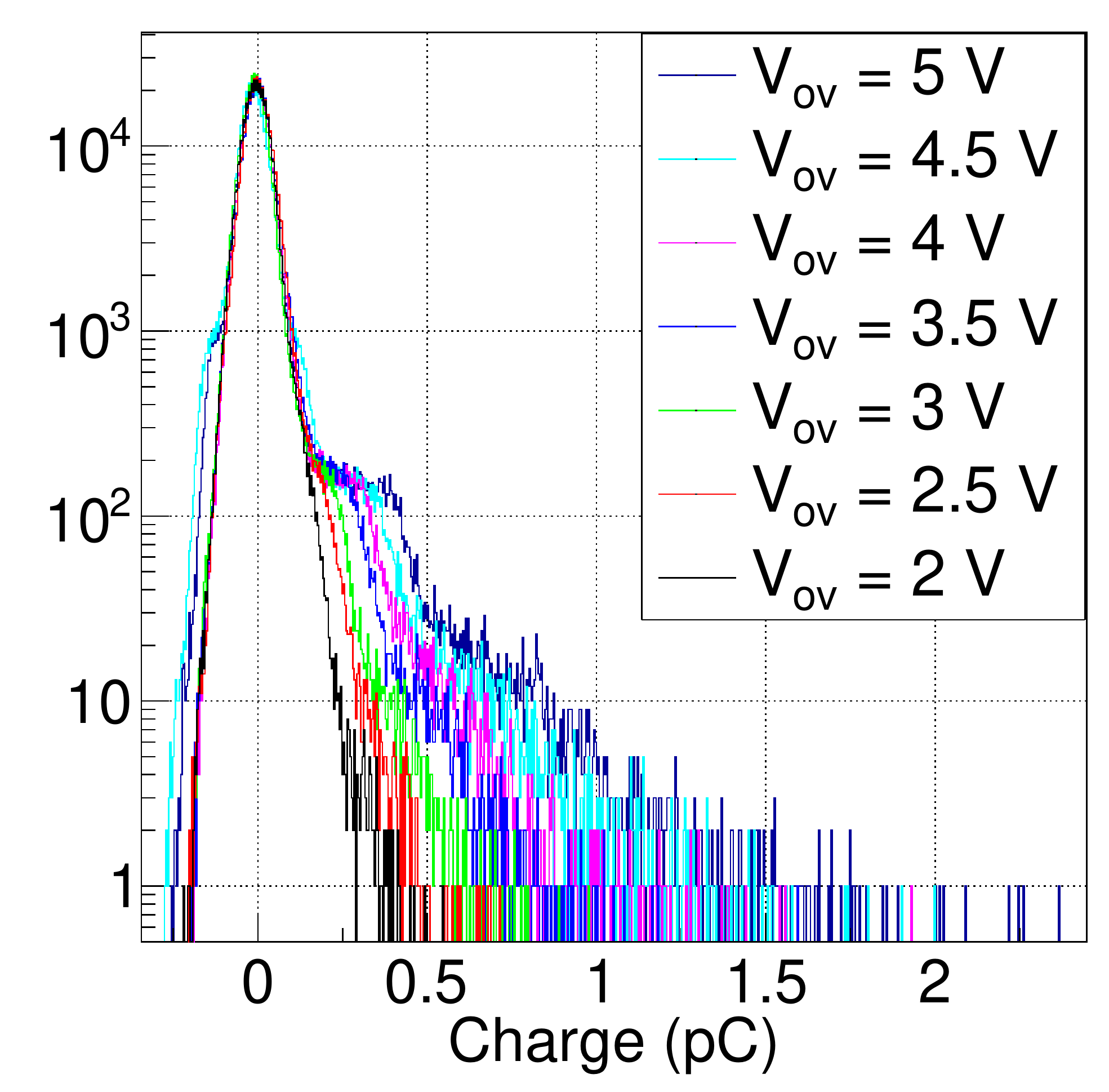}
\caption{\label{fig:noise_chrg_OV}Integrated charge for one of the SiPM channels at different $V_{ov}$.}
\end{minipage}\hspace{1.25cm}%
\begin{minipage}{16pc} 
\includegraphics[height=5.3cm, width=5.75cm]{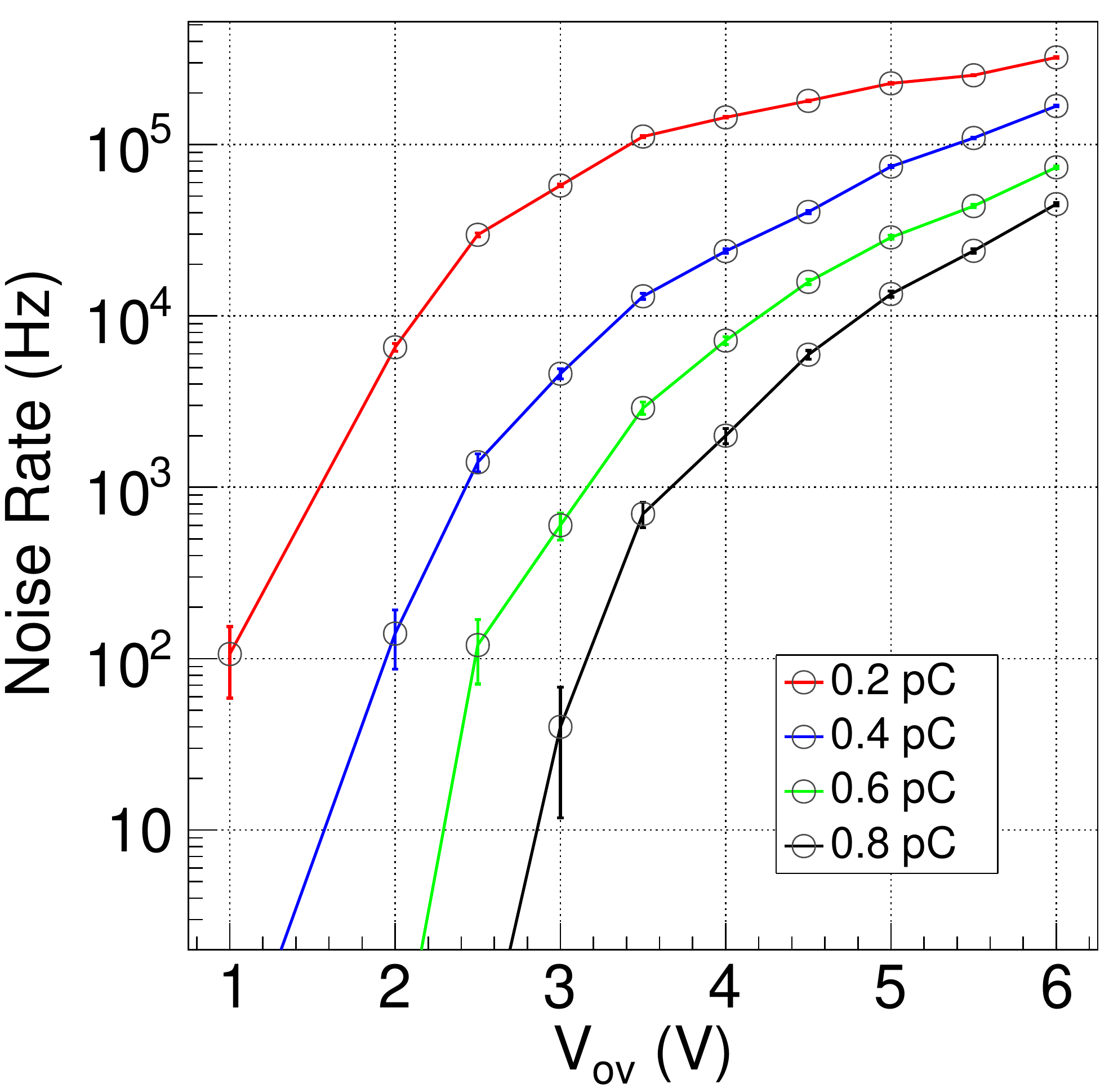}
\caption{\label{fig:noise_sipm} The noise rate for one of the SiPM's as a function of $V_{ov}$ and $q_{th}$.}
\end{minipage}
\end{figure}

The Fig.~\ref{fig:noise_chrg_OV} shows the integrated charge for a random window of 100\,ns for different applied voltages. To avoid any baseline shift, the integrated charge of previous 100\,ns is subtracted from the main signal. This is the main reason for having a bump on the negative side. There is a core Gaussian distribution whose width is almost independent of $V_{ov}$ (0.037\,pC at 2\,V and 0.039\,pC at 6\,V). The source of this noise is mainly due to the readout electronics. The tail of the distribution increases with the applied $V_{ov}$ which is the characteristic of the SiPM and is the source of noise rate in SiPM.

SiPM noise rate as a function of $V_{ov}$ and charge threshold ($q_{th}$) value is shown in Fig.~\ref{fig:noise_sipm}. Along with the cosmic muon signal, this noise rate decides the veto criteria during the experiment.

\section{Muon signal and Noise rate at CMV}
\label{optimization}
The average flux of muons produced by the cosmic rays at sea level is $\sim$1\,muon/cm$^2$/min. The trigger rate is low due to small area times solid angle coverage in the experimental setup is shown in Fig.~\ref{fig:blackbox}. An overnight run is taken to collect a significant amount of data corresponding to each of the $V_{ov}$ values. Both 1\,cm and 2\,cm thick extruded scintillator were used for the study with 5\,m long WLS, which is nearly the same length as the final detector setup. The results discussed in this particular section are from ($60\,cm \times 5\,cm \times 1\,cm$) extruded scintillator. Data is collected with the different relative configurations of fibre and scintillator such that cosmic muon signals are collected for different lengths of WLS fibre, starting from 25\,cm to 475\,cm, but muon always passed through the centre of the same 60\,cm long extruded scintillator.
\subsection{Muon signal in a SiPM}
\begin{figure} [htbp]  
\centering
\includegraphics[height=9.5cm,width=14cm]{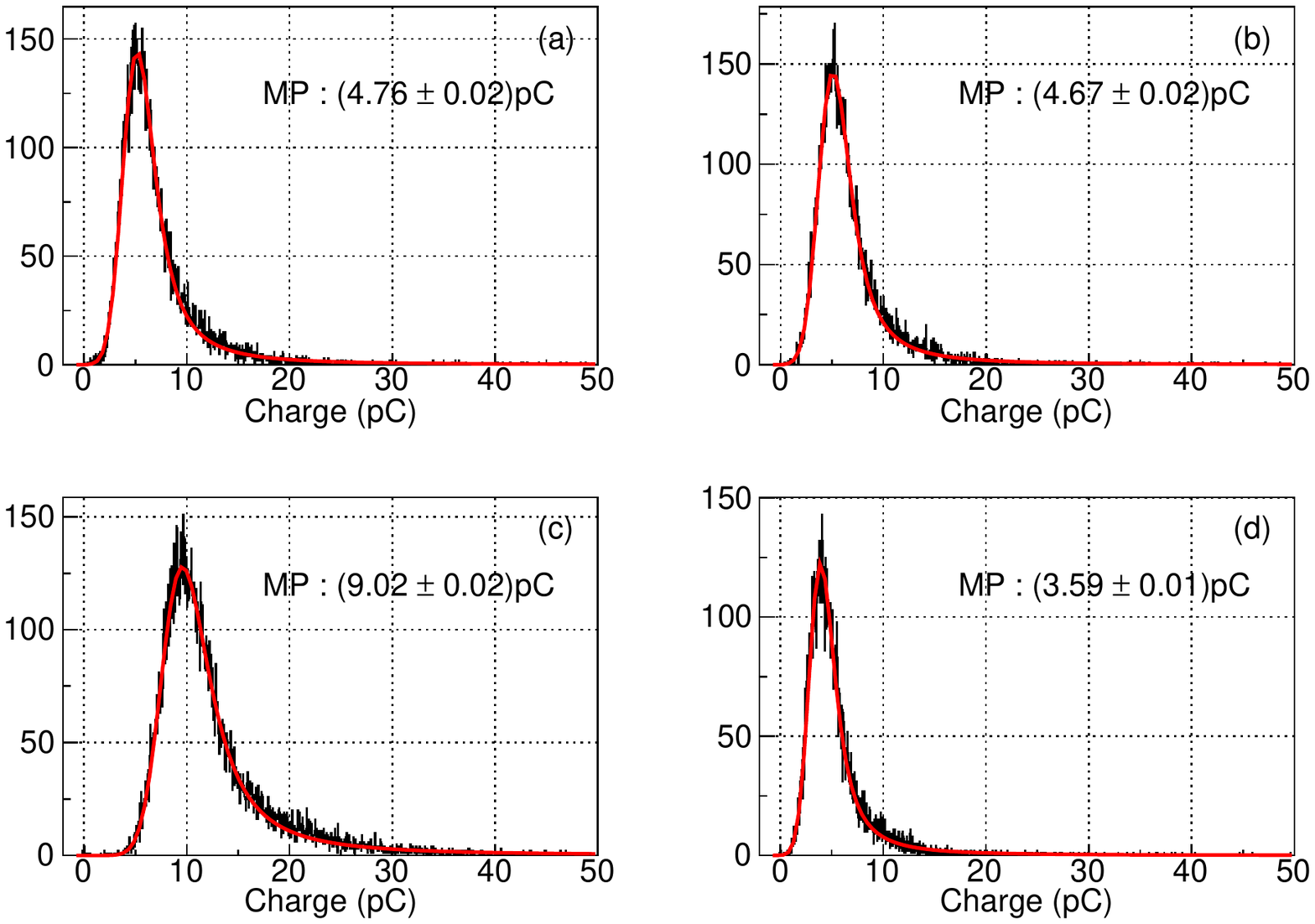}
\caption{Charge distribution for four configurations at  $V_{ov}$ = 3\,V, (a) and (b) are charge distributions for two SiPMs corresponding to same fibre when the readout on both sides are $\sim$ 2.5\,m from the muon incidence position, (c) and (d) are charge distributions for two SiPMs corresponding to same fibre when the readout on one side is $\sim$ 26\,cm and readout on the other side is $\sim$ 4.7\,m from the muon incidence position.}
\label{fig:chargedis}
\end{figure}

\begin{figure} [htbp]
\captionsetup[subfigure]{labelformat=empty}
\centering
\hspace*{-0.5cm}
\begin{subfigure}{0.333\textwidth}
\centering
\includegraphics[height=5cm,width=4.85cm]{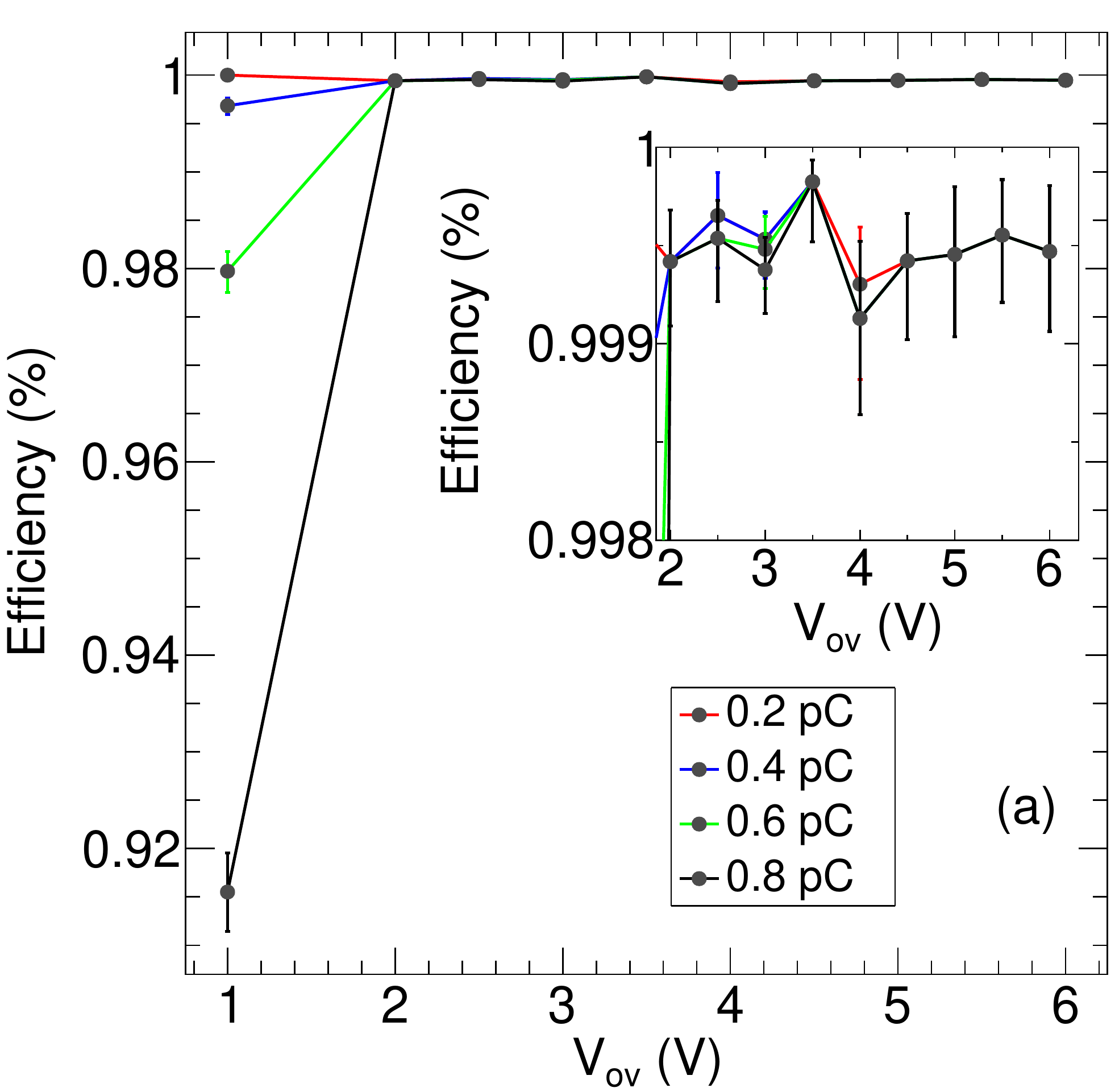}
\caption{} 
\label{fig:near}
\end{subfigure}
\begin{subfigure}{0.333\textwidth}
\centering
\includegraphics[height=5cm,width=4.85cm]{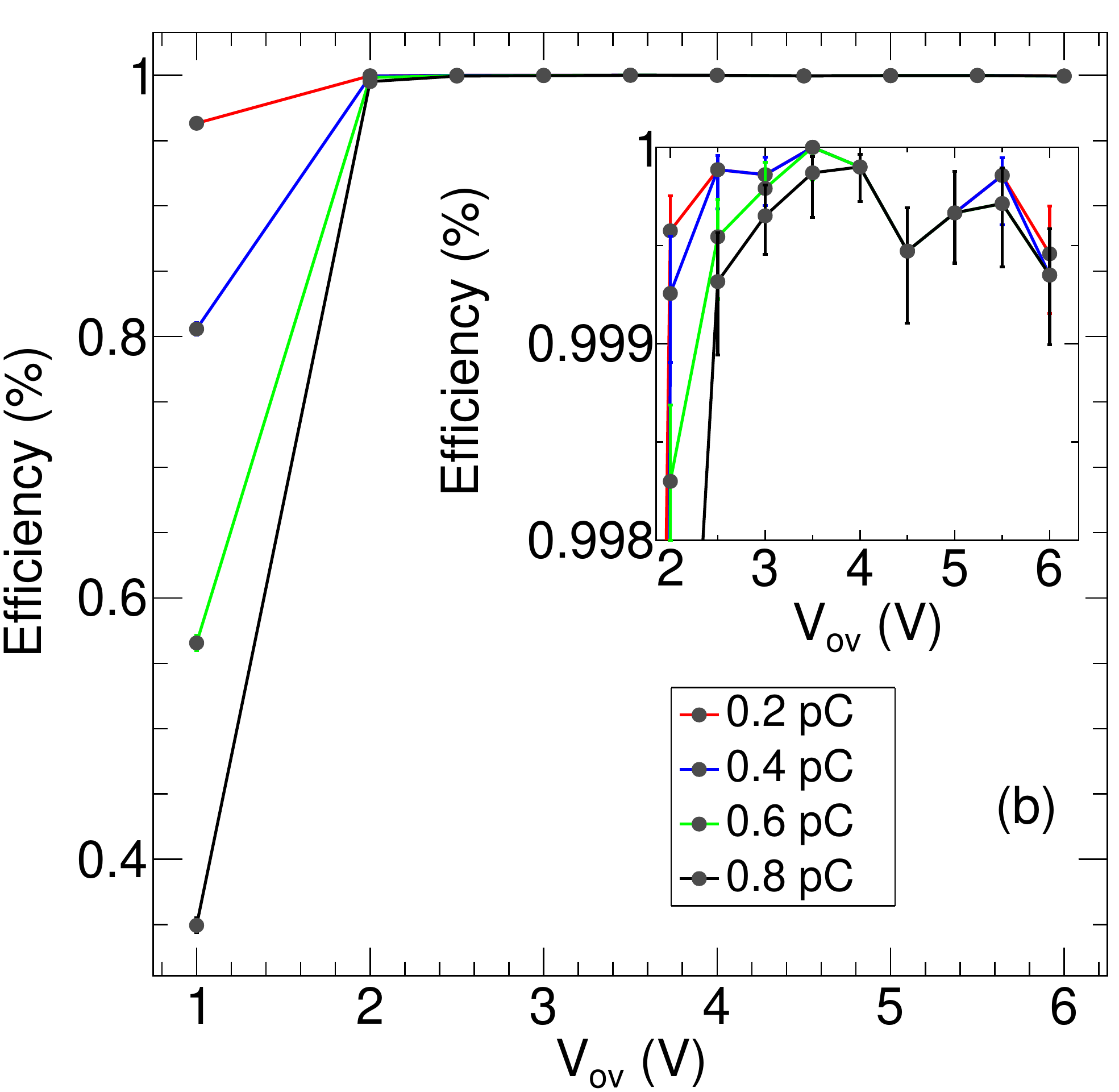}
\caption{}
\label{fig:middle}
\end{subfigure}
\begin{subfigure}{0.333\textwidth}
\centering
\includegraphics[height=5cm,width=4.85cm]{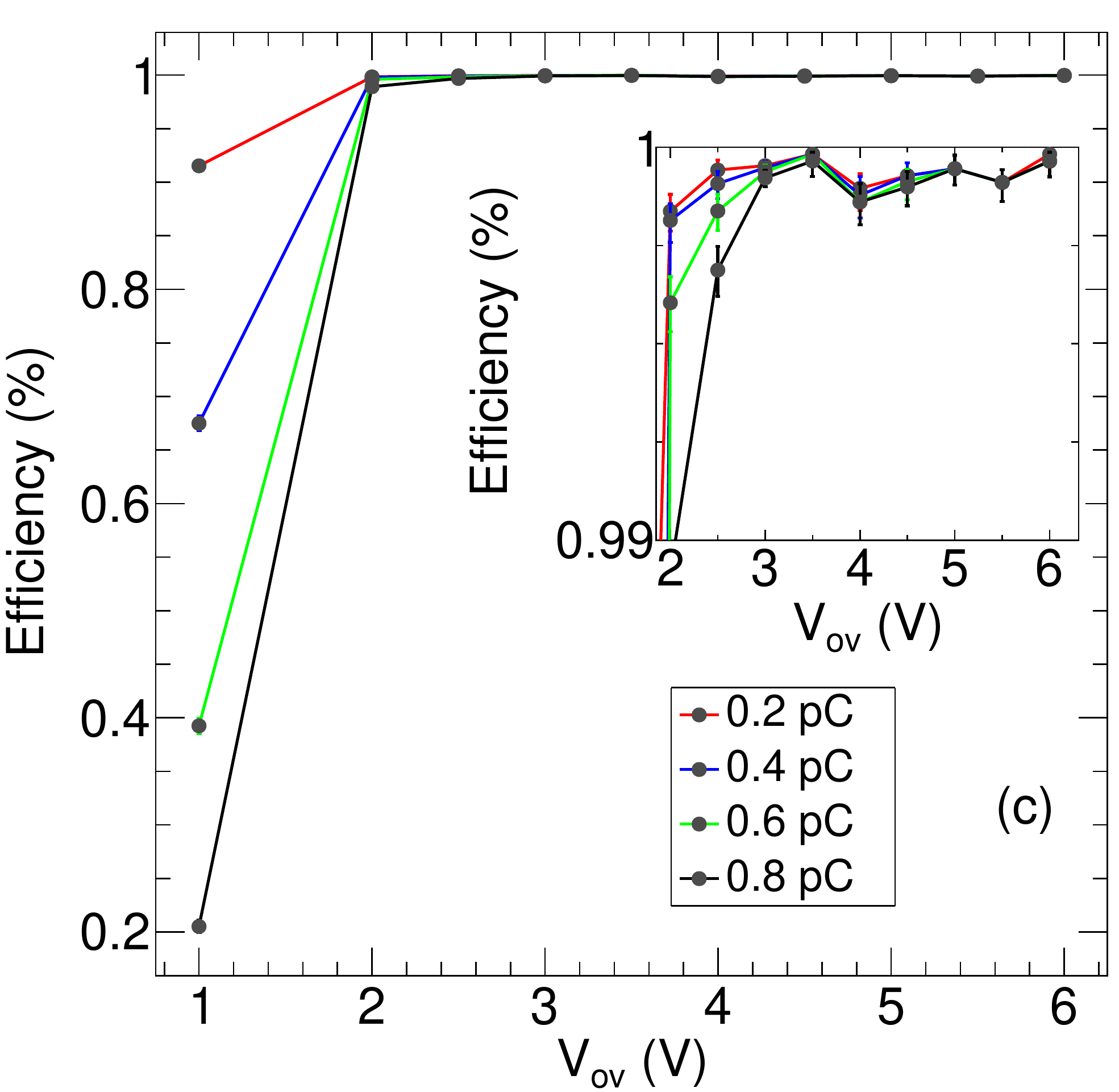}
\caption{}
\label{fig:edge}
\end{subfigure}
\caption{Muon detection efficiency as a function of $V_{ov}$ and $q_{th}$ for one of the SiPM channel on extruded scintillator when the muon position is (a) $\sim$ 27.5\,cm away from readout (b) $\sim$ 2.5\,m away from readout and (c) $\sim$ 4.75\,m away from readout.}
\label{fig:effi_sipm}
\end{figure}

Depending on the energy deposited by the cosmic muon in the extruded scintillator, a certain number of scintillating photons are produced, propagated, absorbed, travel in the WLS fibre, guided towards the SiPM surface, and thus produced the signal. The charge distributions for four configurations at $V_{ov}$ = 3\,V are shown in Fig.~\ref{fig:chargedis}. The expected signal is low due to the attenuation of light when the photon propagates a long distance inside the fibre. All these distributions are fitted with the Gaussian convoluted Landau function for the muon signal and another Gaussian function for the pedestal. The gain of single photoelectron at $V_{ov}$ = 3\,V is $\sim$0.215\,pC. So, the average number of photoelectrons detected corresponding to the shortest and longest path travelled in the fibre are (41.95 $\pm$ 0.09) and (16.70 $\pm$ 0.04) respectively.\\
Muon detection efficiency of these four configurations as a function of different values of $q_{th}$ and $V_{ov}$ are shown in Fig.~\ref{fig:effi_sipm}\,\footnote{Lines joining two points are meant to guide the eye. Inset plots are the same points in reduced range. The same conventions are followed in the next few plots.}. Efficiency reduces with the increase of $q_{th}$ and decrease of $V_{ov}$ and minimum efficiency for Fig.~\ref{fig:edge}, when the photon travels the longest path in the fibre.

\subsection{Cosmic muon detection efficiency in CMV}
\begin{figure} [htbp]  
\centering
\includegraphics[height=3.5cm,width=12.99cm]{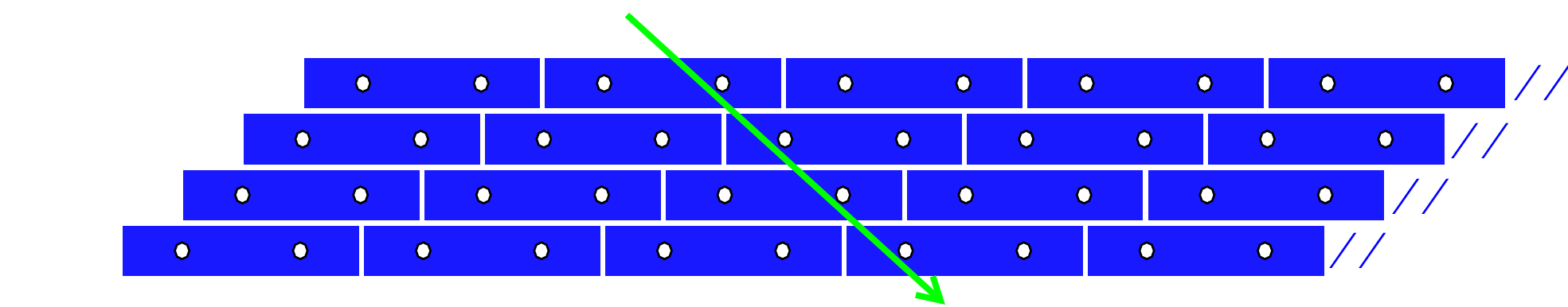}
\caption{A section of top veto detector. Each hole in a extruded scintillator represents the fibre position as well as SiPM mounting position.}
\label{fig:stagger}
\end{figure}

The configuration of the top veto detector is shown in Fig.~\ref{fig:stagger} (four staggered scintillator layers along with fibre positions and gaps). There is a finite gap between the two extruded scintillators and also an inactive region due to the light reflective coating, mechanical tolerance, and packing materials. The estimated average effective gap between two scintillators is 2\,mm. So, there is a probability that some muons might pass through these cracks in the veto detector system, thus contributing to the inefficiency of the veto detector.

\begin{figure} [htbp]
\captionsetup[subfigure]{labelformat=empty}
\centering
\hspace*{-0.5cm}
\begin{subfigure}{0.333\textwidth}
\centering
\includegraphics[height=5cm,width=4.85cm]{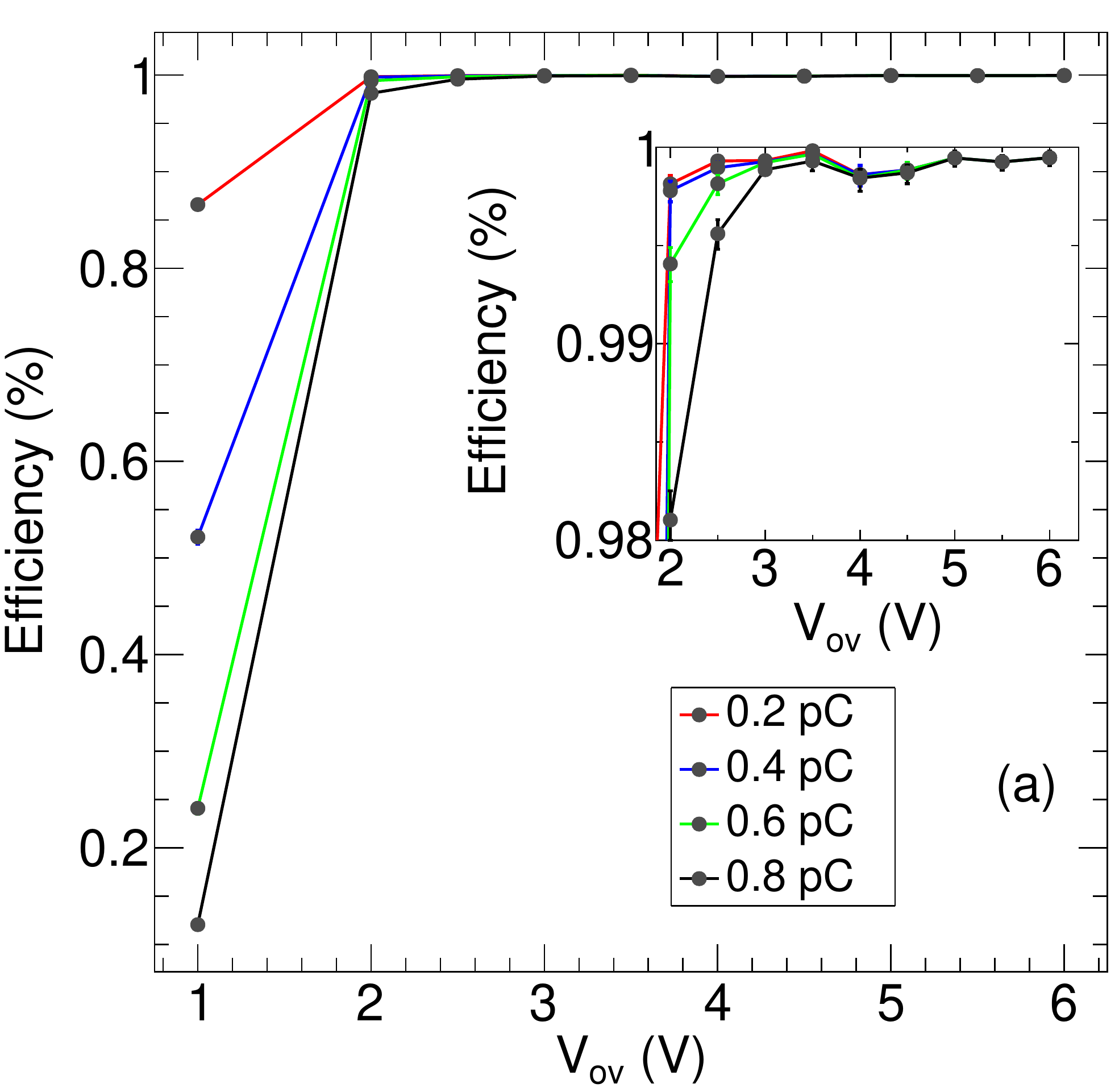}
\caption{}
\label{fig:edgeall4}
\end{subfigure}
\begin{subfigure}{0.333\textwidth}
\centering
\includegraphics[height=5cm,width=4.85cm]{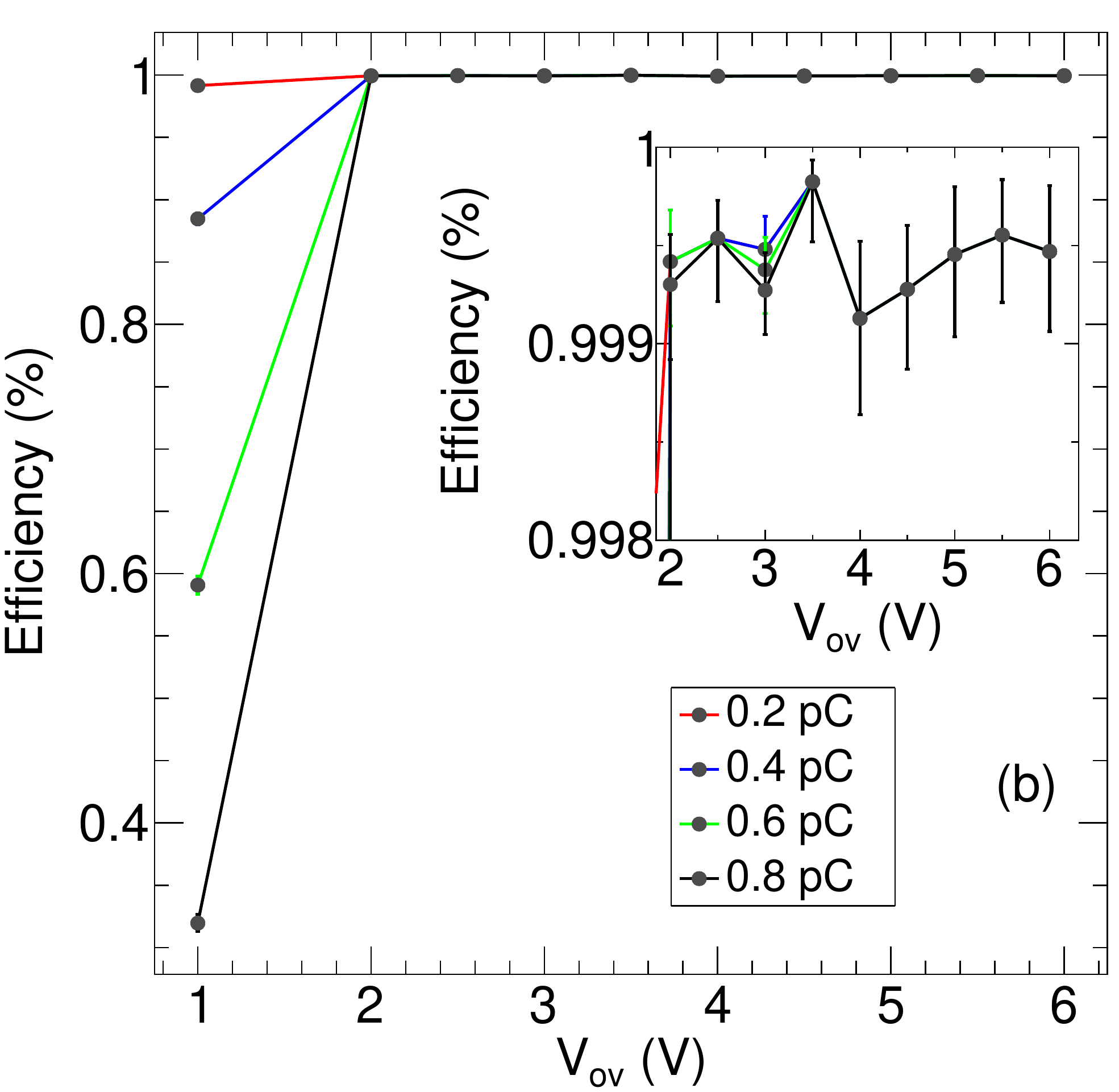}
\caption{}
\label{fig:edgeatl3}
\end{subfigure}
\begin{subfigure}{0.333\textwidth}
\centering
\includegraphics[height=5cm,width=4.85cm]{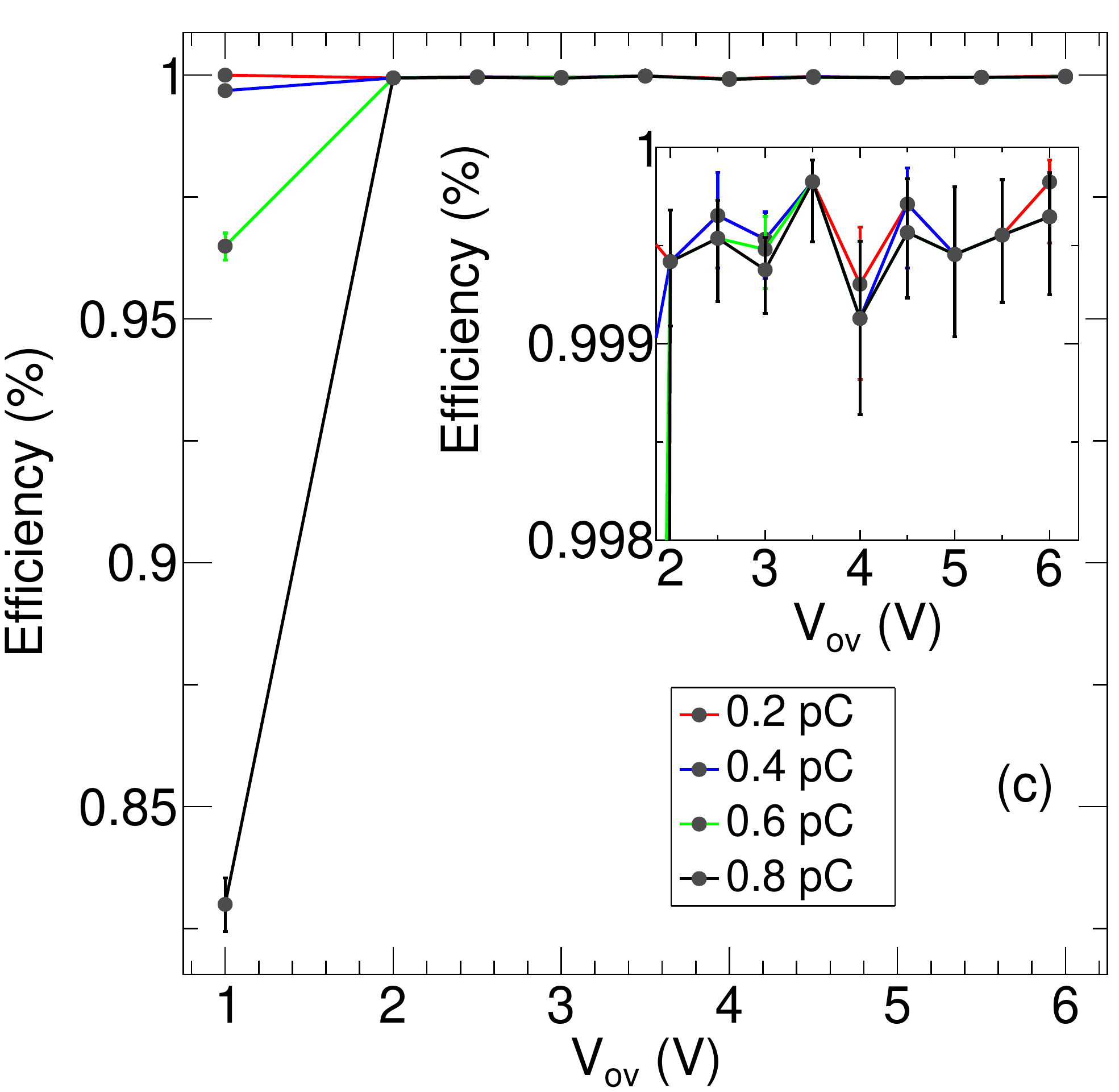}
\caption{}
\label{fig:edgeatl2}
\end{subfigure}
\caption{The muon detection efficiency as a function of $V_{ov}$ and $q_{th}$ when readout on one side is $\sim$ 4.75\,m and other side is 27.5\,cm away from muon incidence position for (a) All four, (b) At least three and (c) At least two SiPM signals are above the threshold.}
\label{fig:edge_scint_effi}
\end{figure}

\begin{figure} [htbp]
\captionsetup[subfigure]{labelformat=empty}
\centering
\hspace*{-0.5cm}
\begin{subfigure}{0.333\textwidth}
\centering
\includegraphics[height=5cm,width=4.85cm]{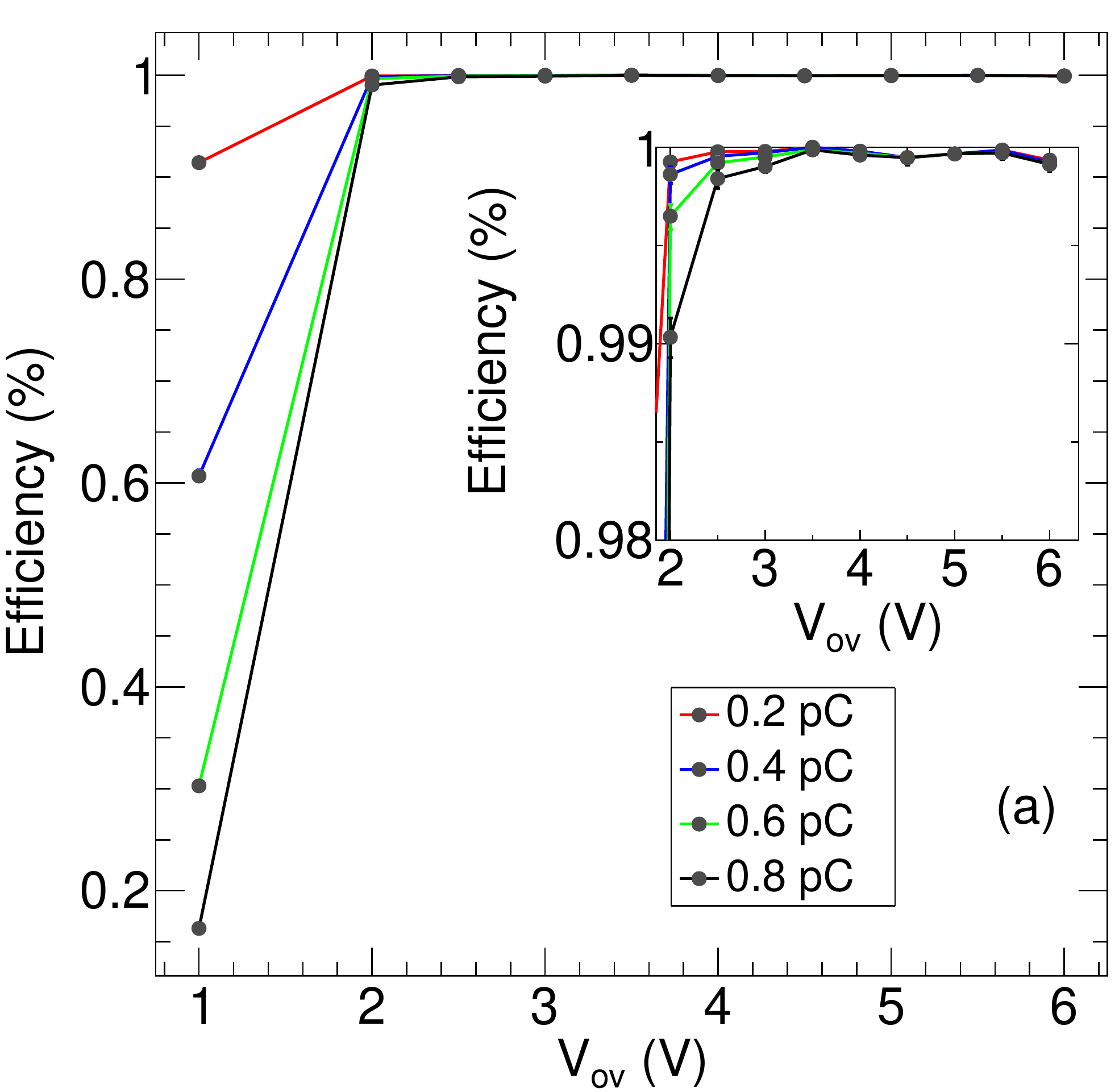}
\caption{}
\label{fig:midall4}
\end{subfigure}
\begin{subfigure}{0.333\textwidth}
\centering
\includegraphics[height=5cm,width=4.85cm]{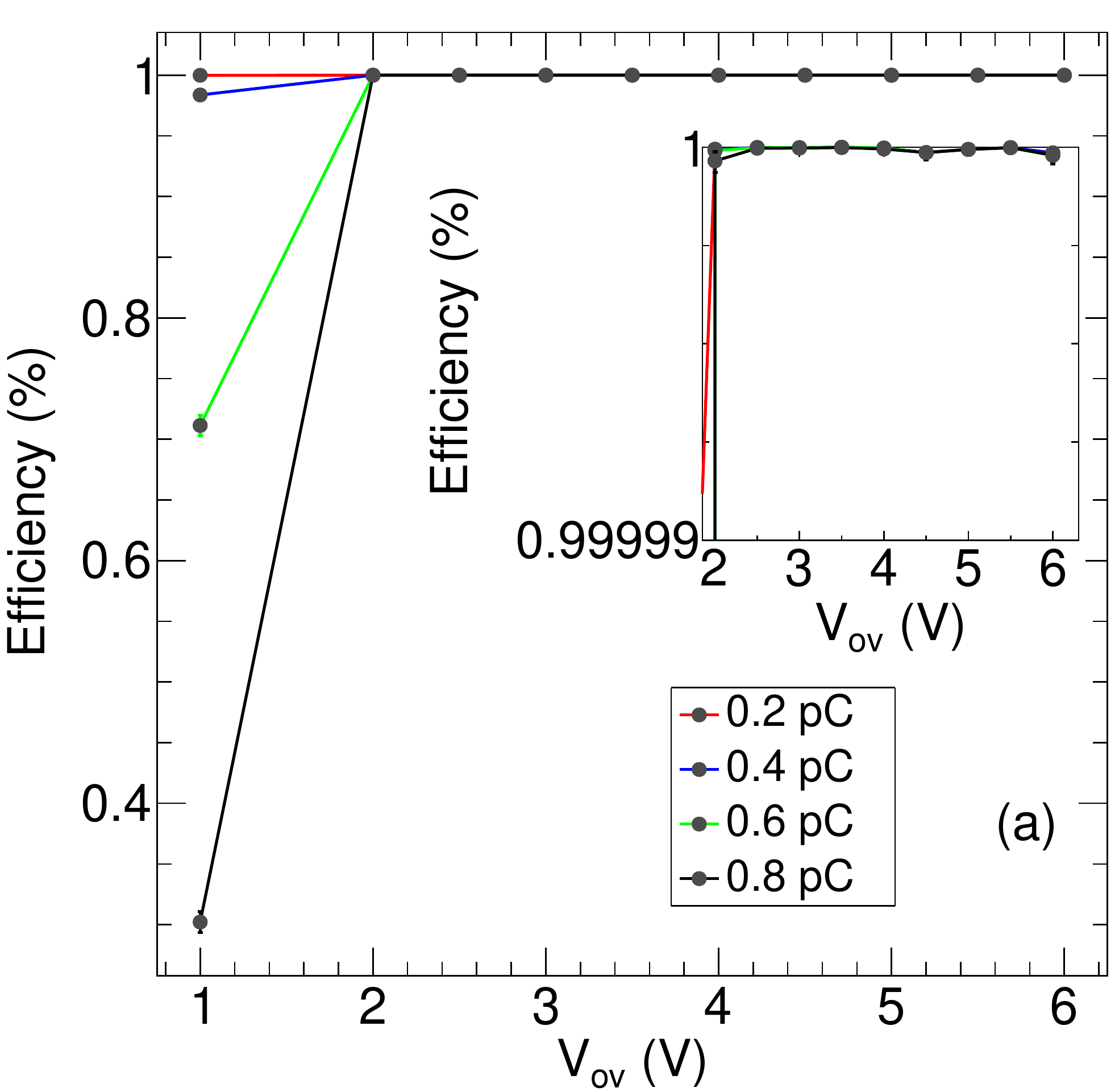}
\caption{}
\label{fig:midatl3}
\end{subfigure}
\begin{subfigure}{0.333\textwidth}
\centering
\includegraphics[height=5cm,width=4.85cm]{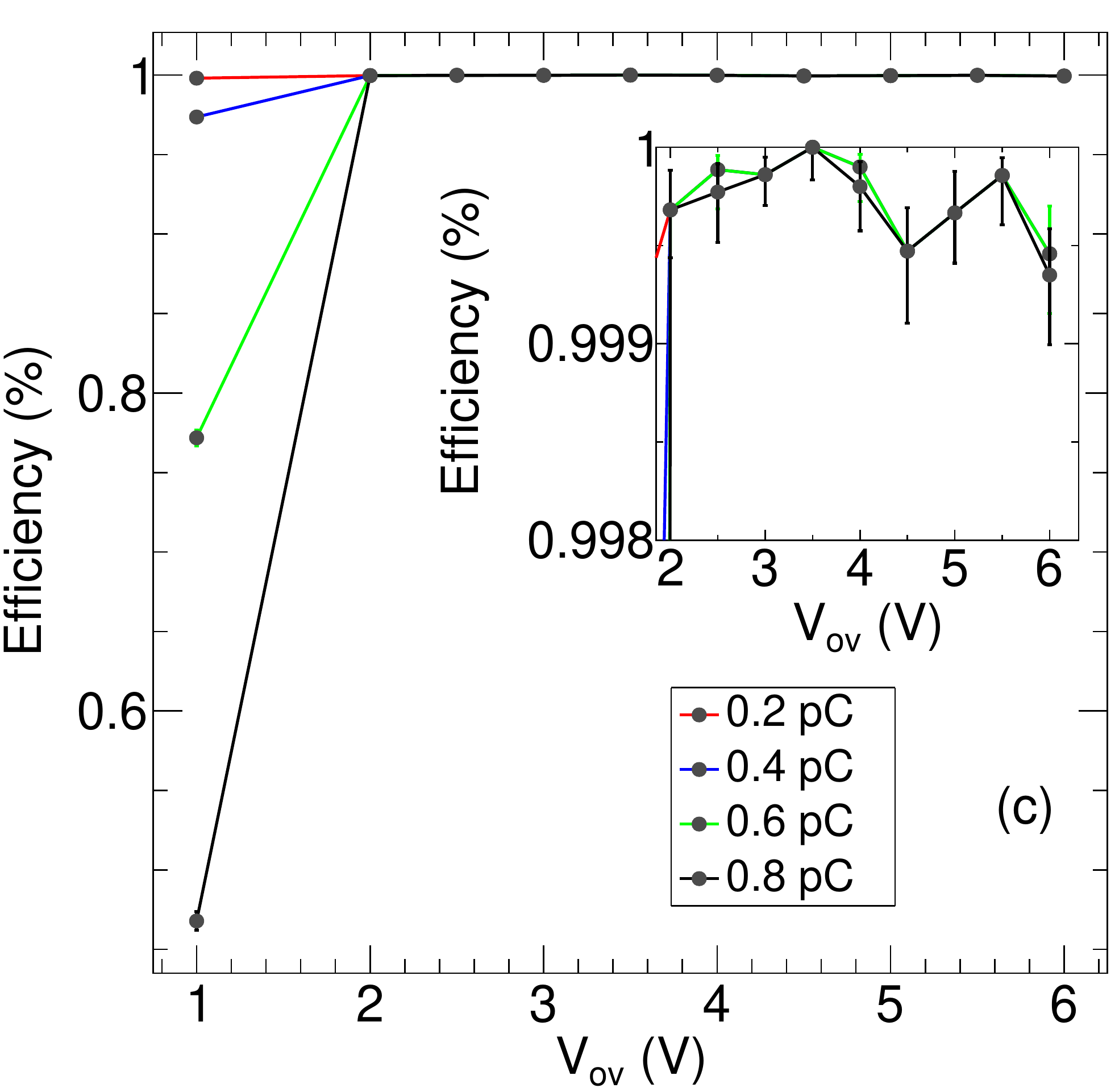}
\caption{}
\label{fig:midatl2}
\end{subfigure}
\caption{The muon detection efficiency as a function of $V_{ov}$ and $q_{th}$ when readout on both sides is $\sim$ 2.5\,m away from muon incidence position for (a) All four, (b) At least three and (c) At least two SiPM signal above the threshold.}
\label{fig:mid_scint_effi}
\end{figure}

A conservative estimate of the average efficiency of individual scintillators when muons pass through the edge of the scintillator resulting in  low  signals from two far end SiPMs due to the attenuation in $\sim$ 5\,m WLS fibre. The efficiency of the scintillator as a function of $V_{ov}$ for various $q_{th}$ thresholds and two, three and four fold coincidences are plotted in Fig.~\ref{fig:edge_scint_effi} and Fig.~\ref{fig:mid_scint_effi}.

\begin{table}[htbp]
\centering
\begin{tabular}{|l|l|l|l|}
\hline
Layer Efficiency  & \multicolumn{3}{c|}{Cosmic muon efficiency ($\%$)}   \\ \hline
& \multicolumn{3}{c|}{Combination of layers for vetoing criteria}\\ \hline
& 2/3 \hspace{1.25cm} & 2/4 \hspace{1.25cm} & 3/4 \\
\hline
0.970 & 99.0580   &  99.9488   &  98.1671   \\
\hline
0.975 & 99.2452   &  99.9653   &  98.5252   \\
\hline
0.980 & 99.4206   &  99.9784   &  98.8628   \\
\hline
0.985 & 99.5839   &  99.9882   &  99.1796   \\
\hline
0.990 & 99.7350   &  99.9949   &  99.4751   \\
\hline
0.995 & 99.8737   &  99.9988   &  99.7487   \\
\hline
\end{tabular}
\caption{Estimated overall CMV efficiencies for various individual layer efficiencies and their coincidence criteria, considering 2\,mm gap between extruded scintillators in a plane. Here 2/3 implies that out of three layers, two of them have signal above threshold and similarly for 2/4 and 3/4 implies that out of four layers, two/three of them have signal above threshold.}
\label{table0}
\end{table}

\begin{figure} [htbp]
\captionsetup[subfigure]{labelformat=empty}
\centering
\begin{subfigure}{0.45\textwidth}
\centering
\includegraphics[height=5cm,width=4.85cm]{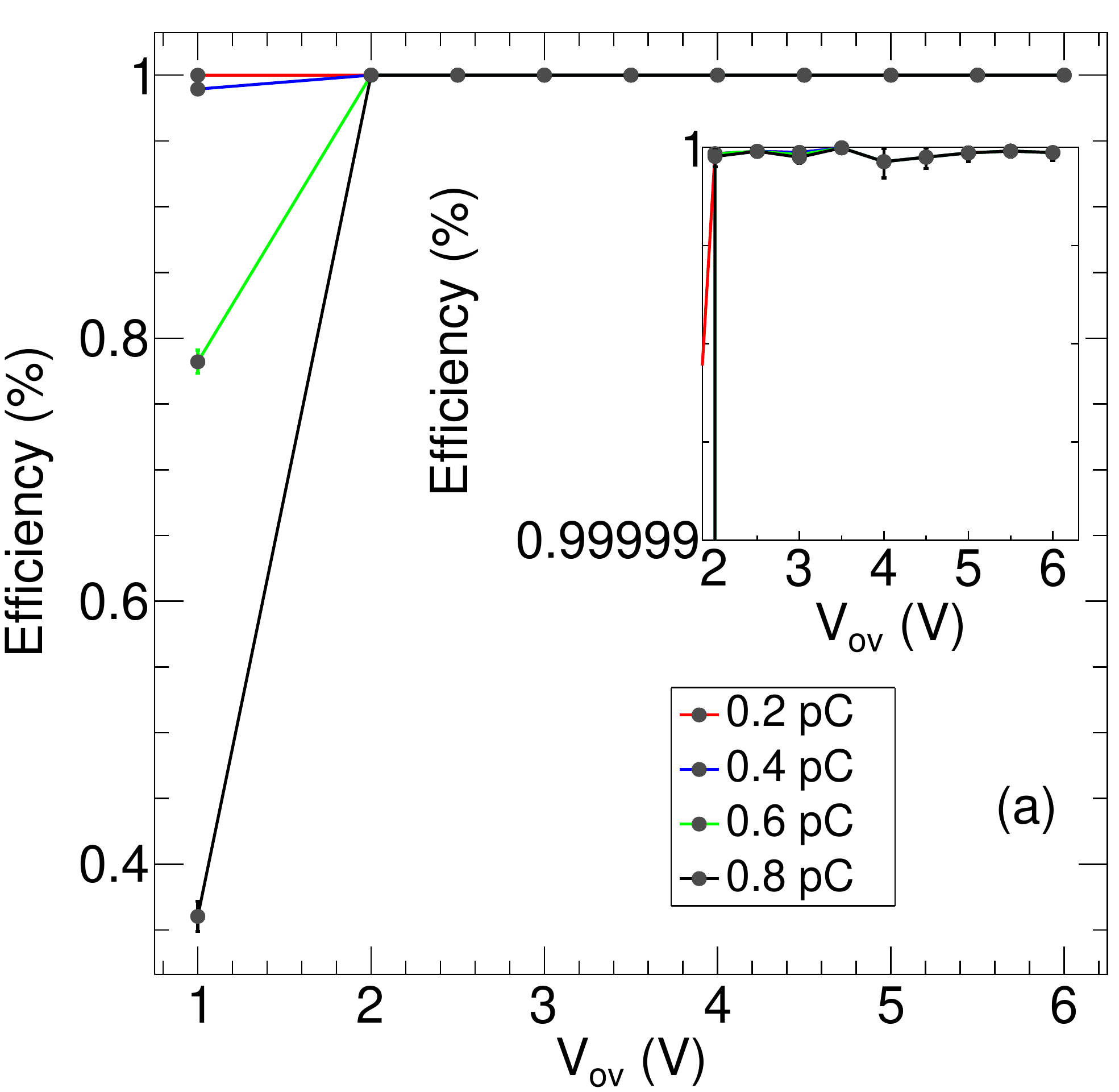}
\caption{}
\label{fig:edge_lay_2in4_atl3}
\end{subfigure}
\begin{subfigure}{0.45\textwidth}
\centering
\includegraphics[height=5cm,width=4.85cm]{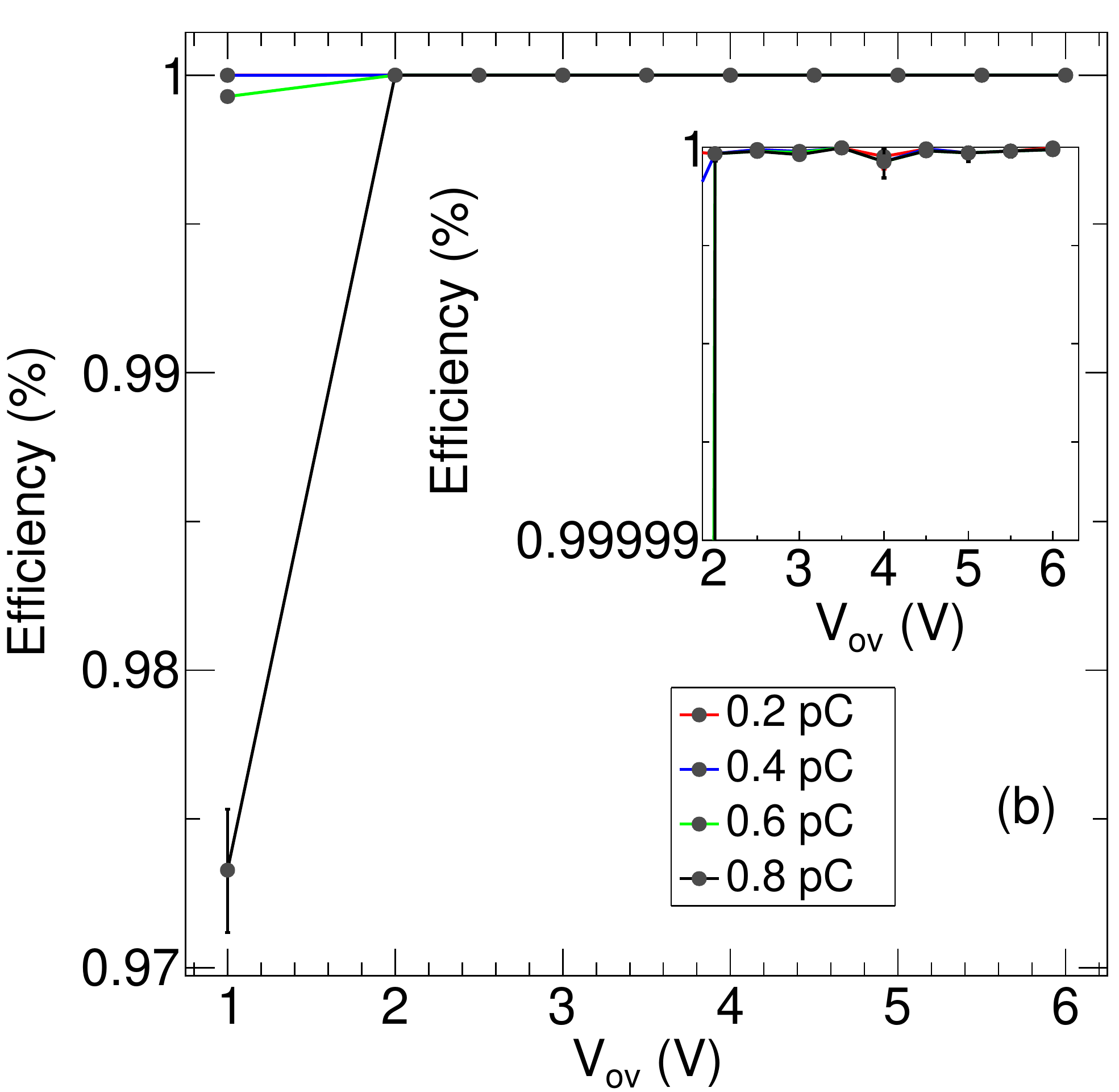}
\caption{}
\label{fig:edge_lay_2in4_atl2}
\end{subfigure}
\caption{The muon detection efficiency as a function of $V_{ov}$ and $q_{th}$ for ``two out of four'' layer criterion when readout on one side is $\sim$ 4.75\,m and other side is 27.5\,cm away from muon coincidence position in a scintillator for (a) at least three SiPM signals and (b) at least two SiPM signals are above the threshold.}
\label{fig:edge_lay_effi_2in4}
\end{figure}

\begin{figure} [htbp]
\captionsetup[subfigure]{labelformat=empty}
\centering
\begin{subfigure}{0.45\textwidth}
\centering
\includegraphics[height=4.85cm,width=4.85cm]{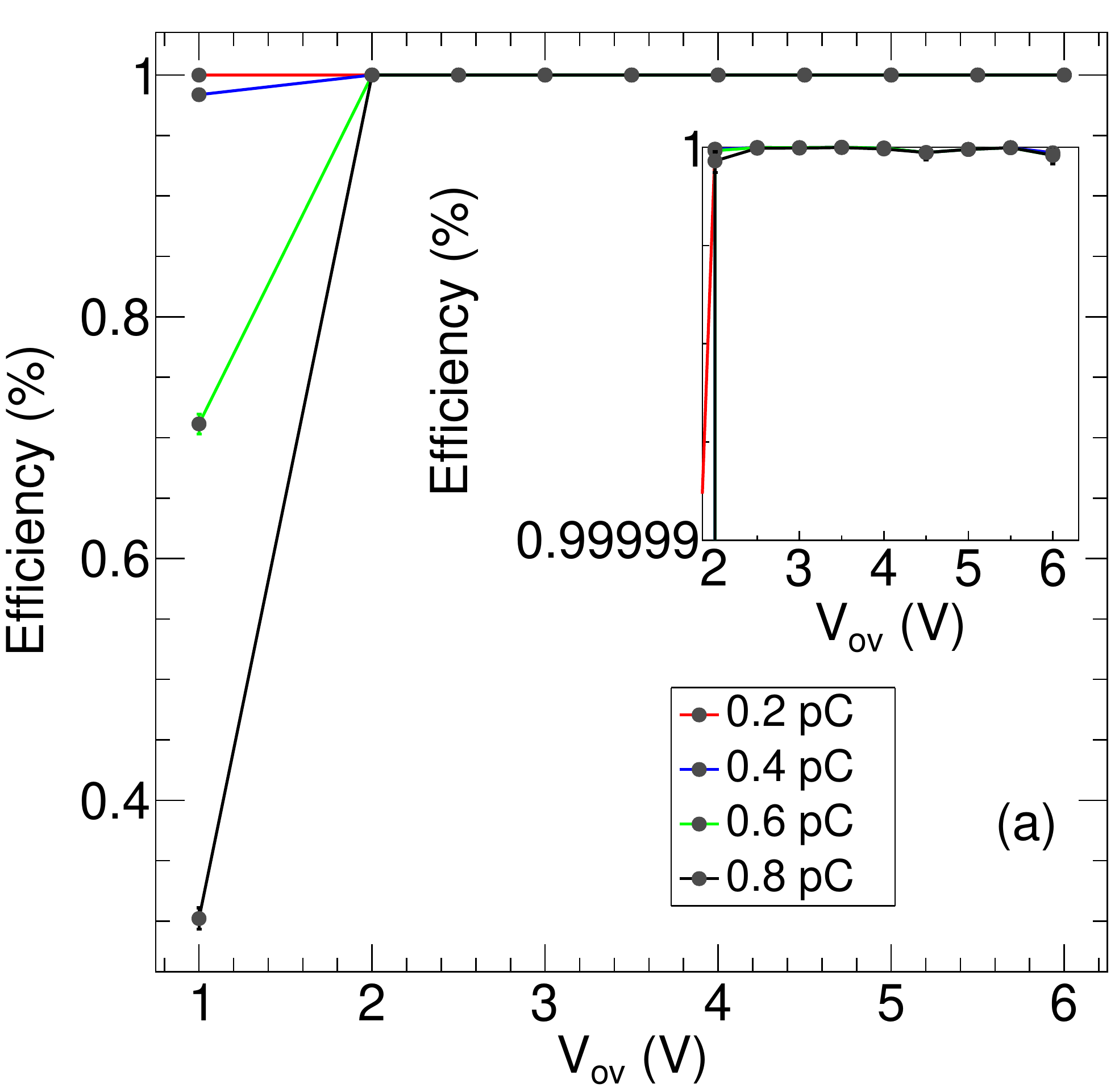}
\caption{}
\label{fig:mid_lay_2in4_atl3}
\end{subfigure}
\begin{subfigure}{0.45\textwidth}
\centering
\includegraphics[height=4.85cm,width=4.85cm]{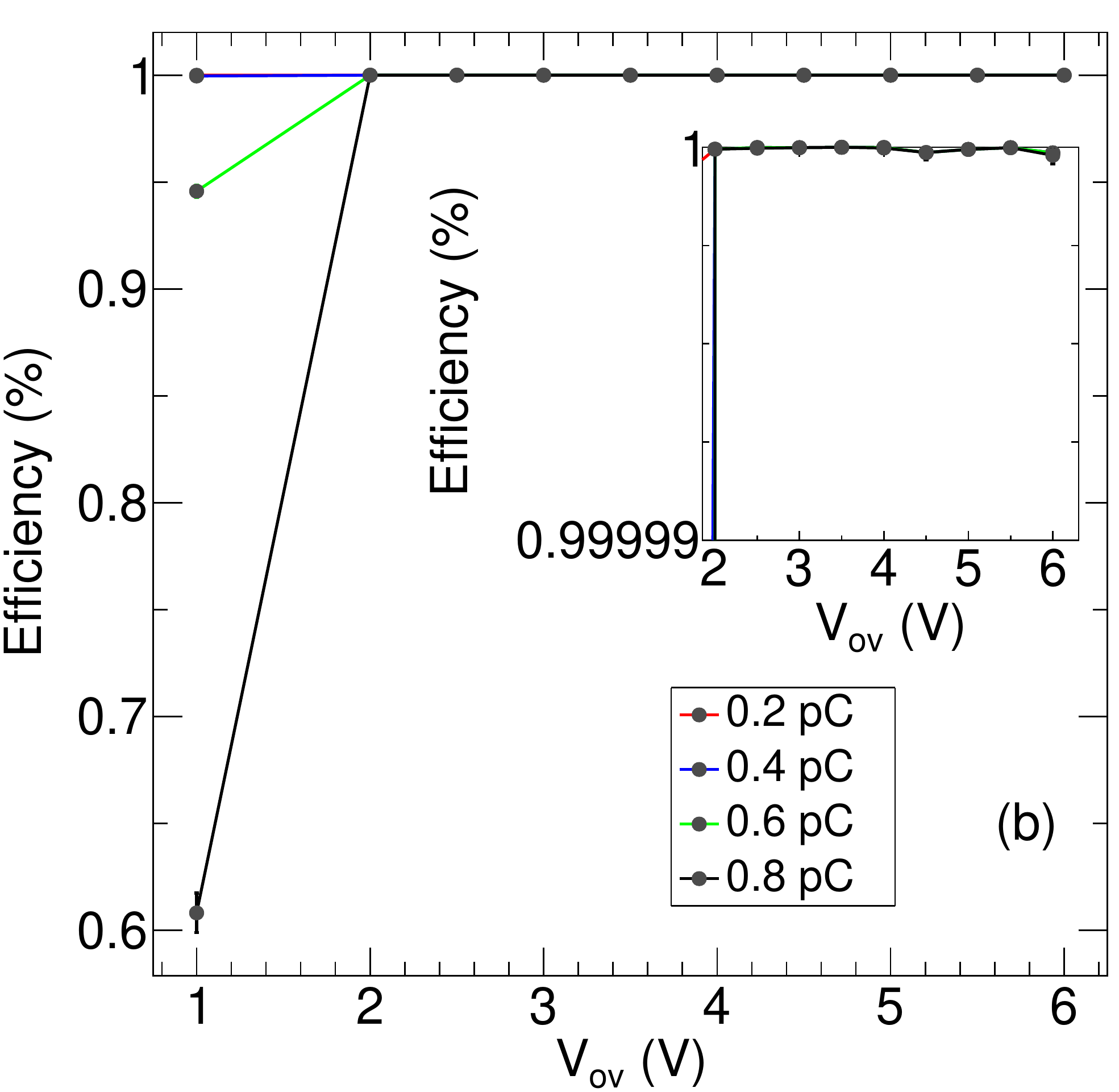}
\caption{}
\label{fig:mid_lay_2in4_atl2}
\end{subfigure}
\caption{The muon detection efficiency as a function of $V_{ov}$ and $q_{th}$ for ``two out of four'' layer criterion when readout on both sides are $\sim$2.5\,m away from muon coincidence position in a scintillator for (a) at least three SiPM signals and (b) at least two SiPM signals are above the threshold.}
\label{fig:mid_lay_effi_2in4}
\end{figure}

Individual scintillator efficiency with four-fold configuration was found to be very poor and one can not consider that as a veto criteria. Also, there is a possibility that a few SiPMs might malfunction during the operation, thus effectively altering the selection criteria. Due to a finite gap between the two scintillators, we can not consider the ``AND'' logic of all scintillators to veto a muon. So, a few possibilities, such as two out of four or three out of four or two out of three scintillator layers may be considered for vetoing the muon signal. Cosmic muon efficiencies for these three possibilities are estimated as a function of layer efficiency considering dead gap of 2\,mm and are shown in Table~\ref{table0} and the required efficiency can be achieved by ``two out of four'' layer trigger criterion. Though the veto efficiencies for all three possibilities are calculated, only the veto efficiency for the ``two out of four'' possibility is discussed as per the efficiency requirements of the detector. The veto efficiency for the ``two out of four'' layer criterion for different values of scintillator detector efficiency is calculated  and are shown in Fig.~\ref{fig:edge_lay_effi_2in4} and Fig.~\ref{fig:mid_lay_effi_2in4} for different configurations.

\subsection{Noise rate in CMV}
\begin{figure} [htbp]
\captionsetup[subfigure]{labelformat=empty}
\centering
\begin{subfigure}{0.49\textwidth}
\includegraphics[height=5.cm,width=6.25cm]{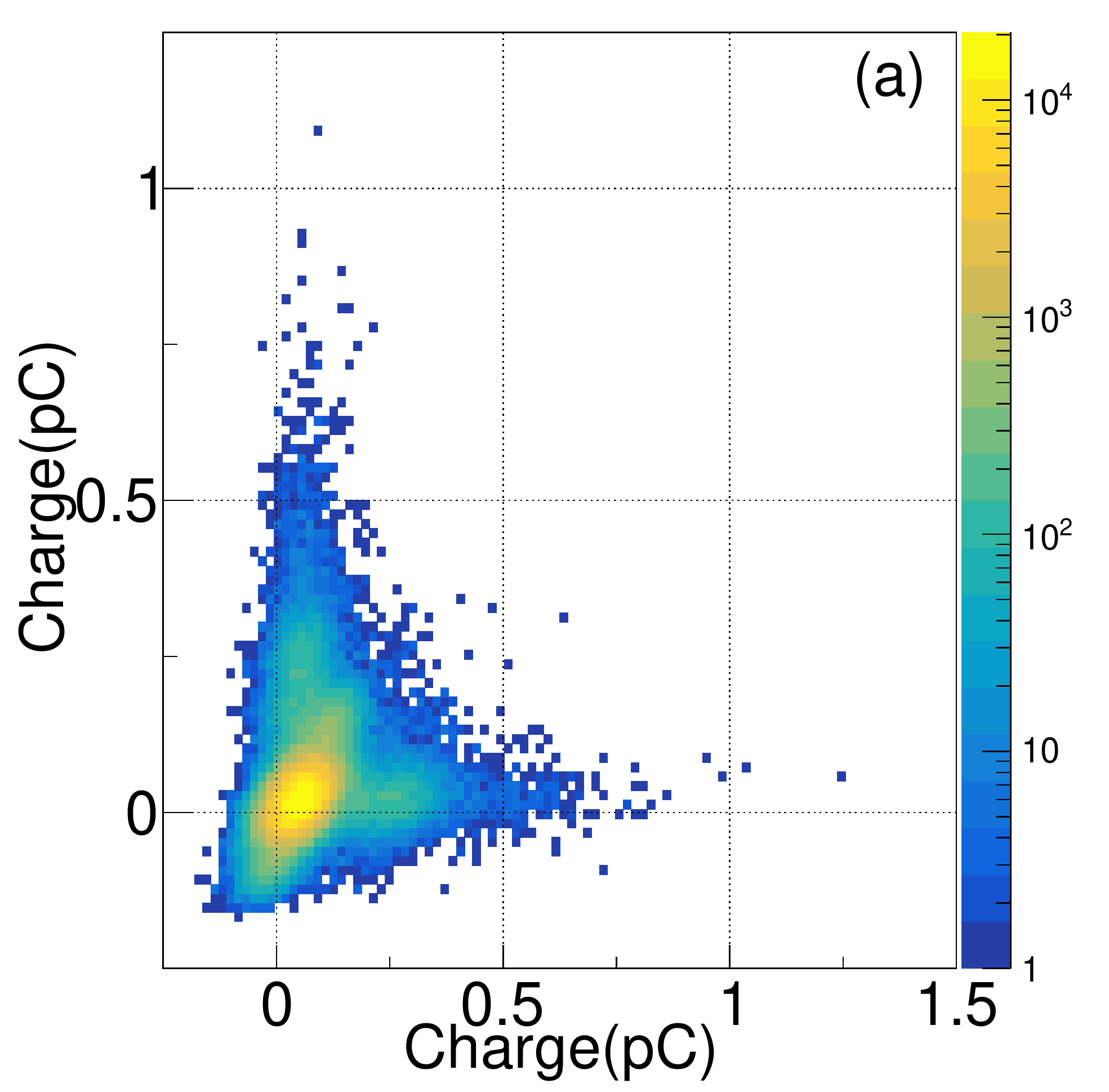}
\caption{}
\label{fig:samefib}
\end{subfigure}
\begin{subfigure}{0.49\textwidth}
\includegraphics[height=5.cm,width=6.25cm]{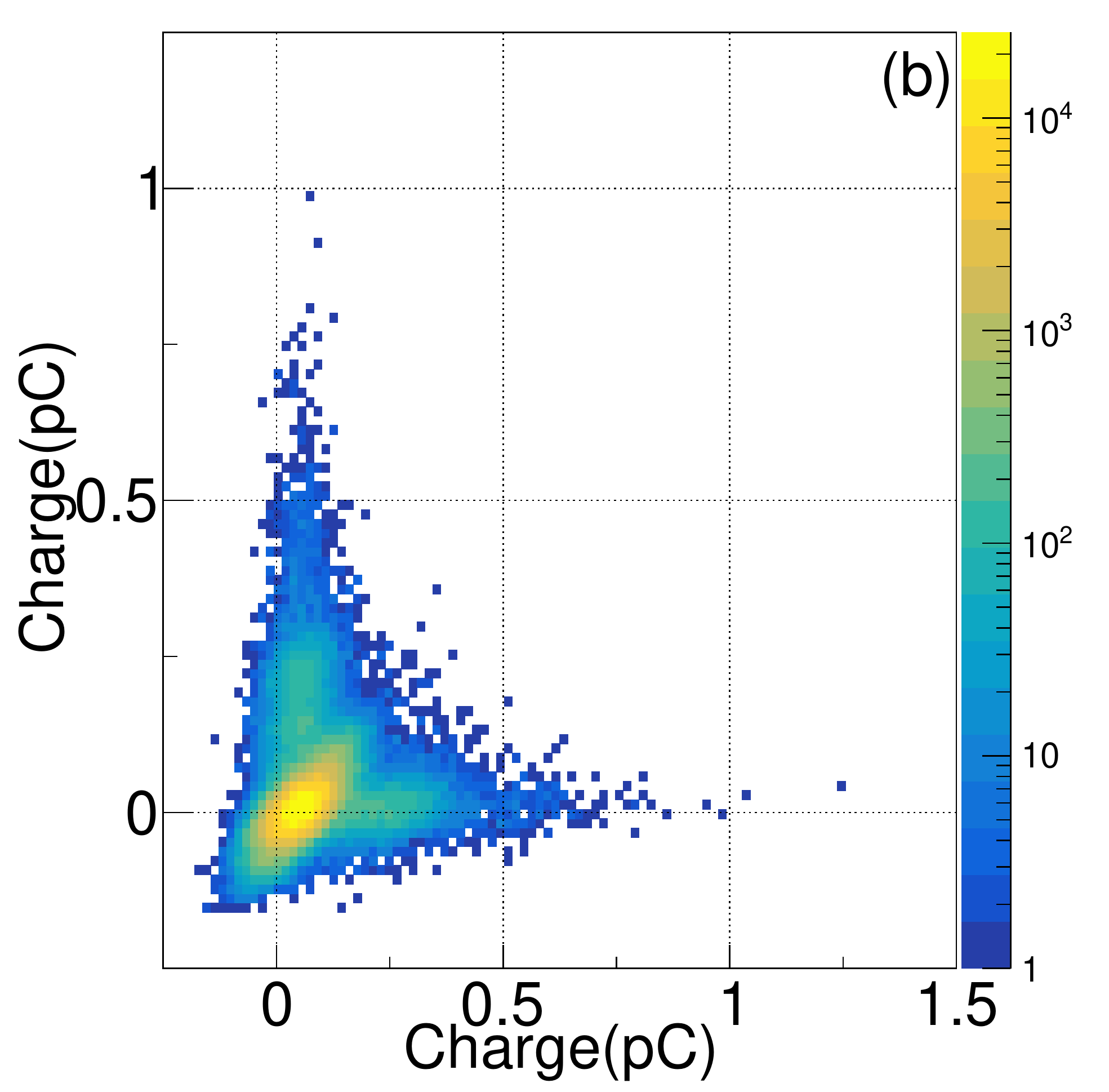}
\caption{}
\label{fig:sameside}
\end{subfigure}
\caption{Correlation between integrated charge of two SiPMs at $V_{ov}$ = 3\,V belonging to the (a) same fibre (b) same side.}
\label{fig:corr2D}
\end{figure}

\begin{figure}[htbp]
\captionsetup[subfigure]{labelformat=empty}
\centering
\begin{subfigure}{0.49\textwidth}
\centering
\includegraphics[height=5.cm,width=6.5cm]{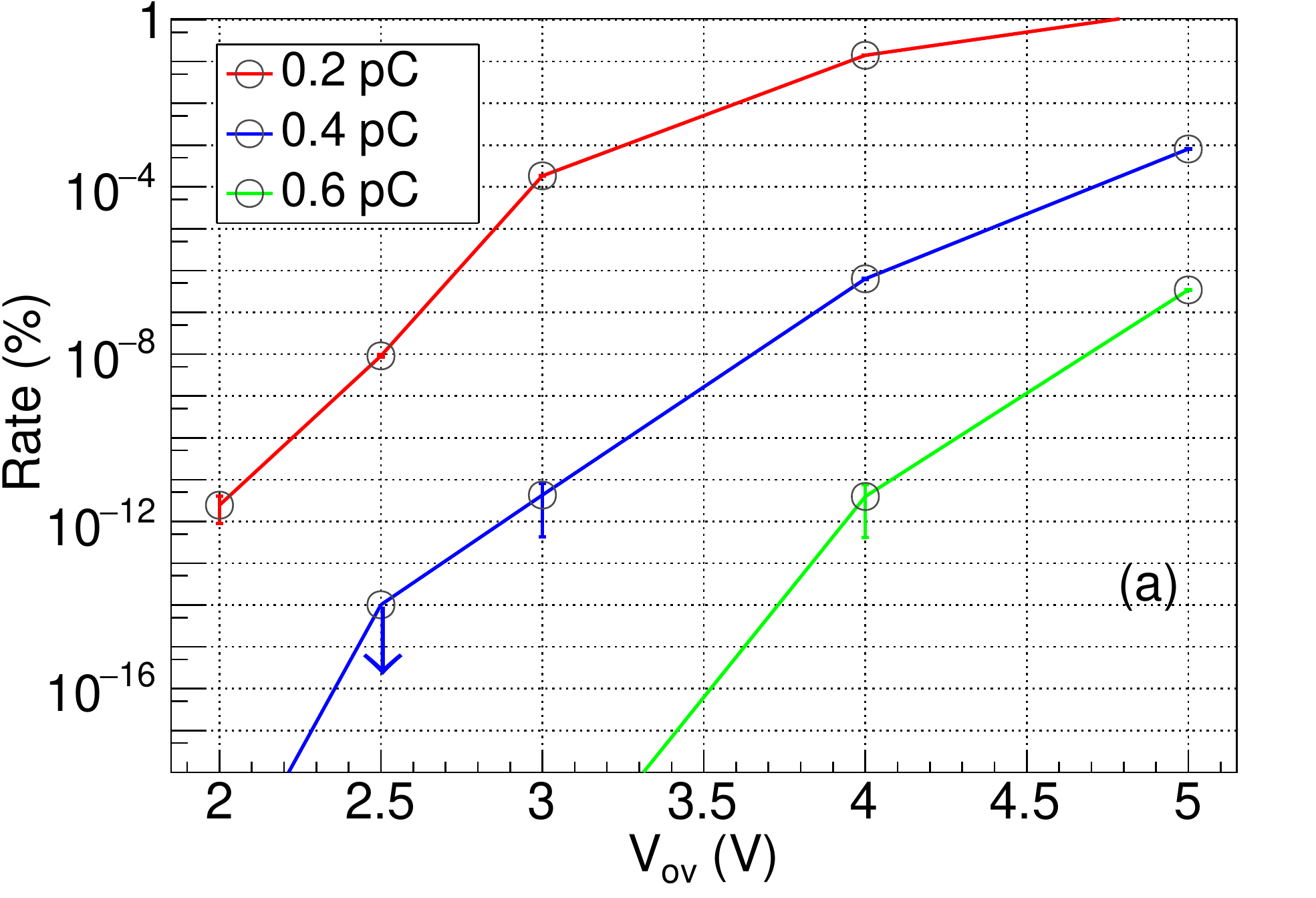}
\caption{}
\label{fig:2O4_2F}
\end{subfigure}
\begin{subfigure}{0.49\textwidth}
\centering
\includegraphics[height=5.cm,width=6.5cm]{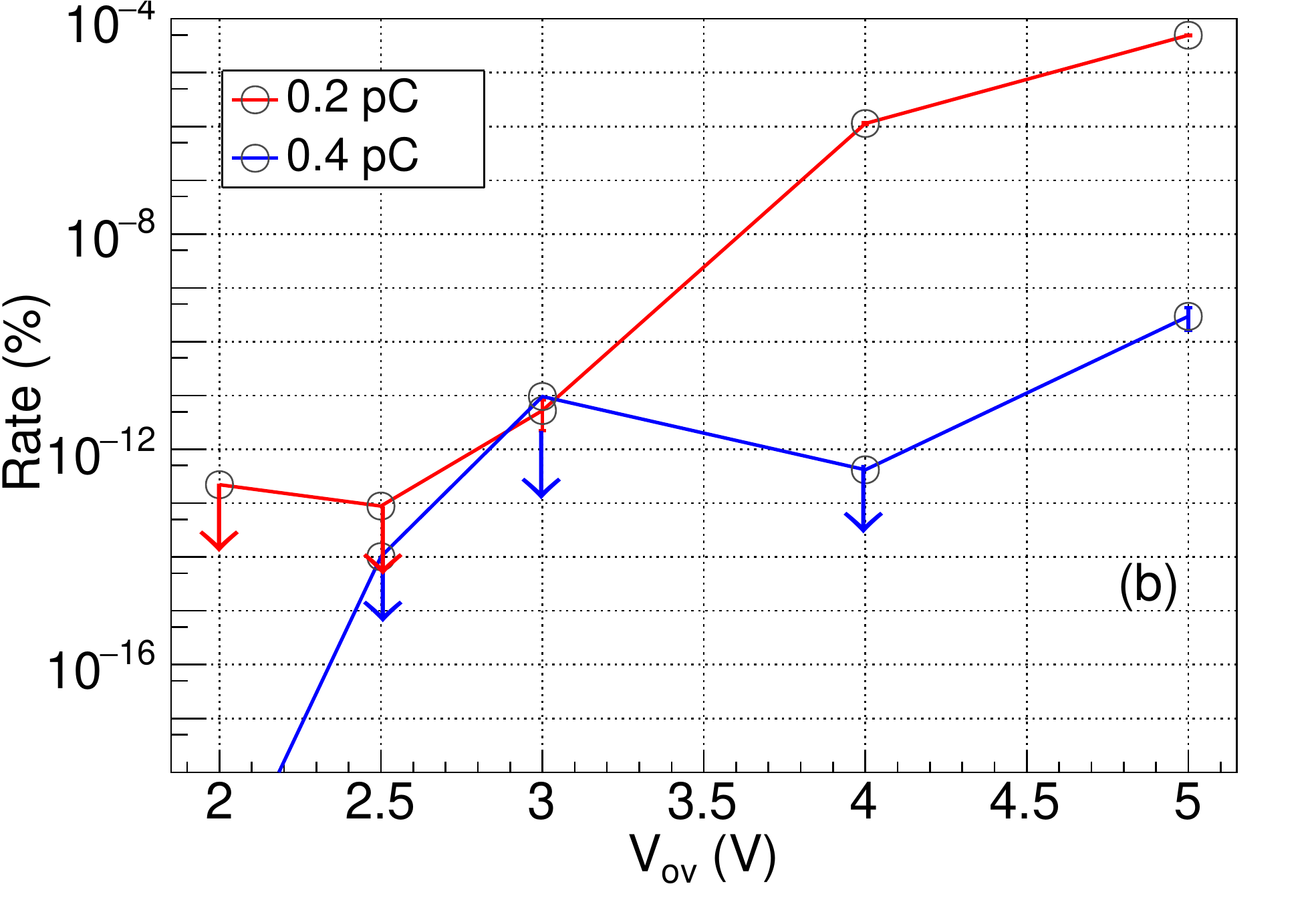}
\caption{}
\label{fig:2O4_3F}
\end{subfigure}
\caption{Random coincidence rate of the CMV detector while 2 out of 4 layers are in coincidence, where random rates in a layer is taken from (a) Table~\ref{table1} and (b) Table~\ref{table2}.}
\label{fig:2O4}
\end{figure}

As shown in Fig.~\ref{fig:corr2D}, during the noise runs, no correlation of noise above 0.2 pC was found between the two SiPMs. Like in the case of cosmic muon triggered events, we have calculated the noise rate of a scintillator while at least two and at least three SiPM signals are above the threshold value as shown in Table~\ref{table1} and Table~\ref{table2} for different $V_{ov}$ using the noise data itself. Due to the very low coincidence rates, a "zero" count was recorded in many configurations. For these cases, a count of 3, which is the upper limit for 95$\%$ Confidence Level (C.L.) was assigned. For arriving at the noise rate of the veto detector, these numbers are propagated and shown as an upper limit. These obtained numbers are consistent with the estimation from uncorrelated single count rate as shown in Fig.~\ref{fig:noise_sipm}. The noise rates for the chance coincidences of ``two out of four'' layers were estimated as shown in Fig.~\ref{fig:2O4} for different $V_{ov}$ and $q_{th}$. 

\begin{table}[htbp]
\centering
\begin{tabular}{|l|l|l|l|}
\hline
V$_{ov}$ (V) & \multicolumn{3}{c|}{Noise Rate $\times$ $10^{-3}$ (Hz)}   \\ \hline
& \multicolumn{3}{c|}{q$_{th}$(pC)}\\ \hline
&  0.2  & 0.4 & 0.6 \\
\hline
2.0 & 1.01 $\pm$ 0.32 & - & - \\
\hline
2.5 & 61.60 $\pm$ 1.99 & \textless 0.07 & - \\
\hline
3.0 & 8766.30 $\pm$ 32.89 & 1.33 $\pm$ 0.94 & - \\
\hline
4.0 &242436.00 $\pm$ 299.65 & 514.58 $\pm$ 8.45 & 1.28 $\pm$ 0.57 \\
\hline
5.0 & 867202.00 $\pm$ 454.39 & 18361.70 $\pm$ 101.00 & 378.7440 $\pm$ 1.56 \\
\hline
\end{tabular}
\caption{Random coincidence rate of any two SiPMs in a scintillator for different $V_{ov}$ and $q_{th}$.}
\label{table1}
\end{table}

\begin{table}[htbp]
\centering
\begin{tabular}{|l|l|l|l|}
\hline
V$_{ov}$ (V) & \multicolumn{3}{c|}{Noise Rate $\times$ $10^{-3}$ (Hz)}   \\ \hline
& \multicolumn{3}{c|}{q$_{th}$(pC)}\\ \hline
&  0.2  & 0.4 & 0.6 \\
\hline
2.0 & \textless  0.30  & - & - \\
\hline
2.5 & \textless  0.19   & \textless  0.07  & - \\
\hline
3.0 & 1.48 $\pm$ 0.43 & \textless  2.00  & - \\
\hline
4.0 & 687.41 $\pm$ 15.96 & \textless  0.42 & \textless  0.77  \\
\hline
5.0 & 4543.10 $\pm$ 32.89 & 11.11 $\pm$ 2.48 & 0.02 $\pm$ 0.01 \\
\hline
\end{tabular}
\caption{Random coincidence rate of any three SiPMs in a scintillator for different $V_{ov}$ and $q_{th}$.}
\label{table2}
\end{table}

\subsection{Optimized overvoltage \texorpdfstring{$V_{ov}$}{Lg} and charge threshold \texorpdfstring{$q_{th}$}{Lg} of SiPM}
From Table~\ref{table6}\color{black},  it is clear that the noise rate for the whole detector system is negligible (\textless$10^{-5}$) for $V_{ov}$ up to 3V. But on the other hand, muon signal is not very large and therefore a large inefficiency is obtained for $V_{ov}$ less than 1.5\,V. Thus, the optimization of $q_{th}$ is tested for the $V_{ov}$ from 2 to 3\,volt. Table~\ref{table6} shows the efficiency for different configurations at $V_{ov}$ = 2, 2.5 and 3\,V along with the chance coincidence rates for these three cases. This table implies the optimum operational scenario is  $V_{ov}$ = 2\,V and $q_{th}$ = 0.4\,pC. This will also be satisfied by the scintillator of thickness 2\,cm.

\begin{table}[htbp]
\centering
\begin{tabular}{|l|l|l|l|l|}
\hline
V$_{ov}$ (V) & \multicolumn{2}{c|}{Efficiency ($\%$)} & \multicolumn{2}{c|}{Noise Rate $\times$ $10^{-12}$ (Hz)}    \\ \hline
& \multicolumn{2}{c|}{q$_{th}$(pC)} & \multicolumn{2}{c|}{q$_{th}$(pC)}\\ \hline
&  0.2  & 0.4 & 0.2 & 0.4 \\
\hline
2.0 & 100.0  &  99.9995 & 2.45 $\pm$ 1.55 & - \\
\hline
2.5 & 100.0  & 100.0 &  9107.71 $\pm$ 5.88 &  \textless 0.010\\
\hline
3.0 & 100.0 & 100.0 & (184.44 $\pm$ 1.38)$\times$ $10^{6}$  &  4.27 $\pm$  1.84 \\  
\hline           
\end{tabular}
\caption{Noise rate and muon efficiency for the whole veto detector system as a function of $V_{ov}$ and $q_{th}$. Error on efficiency is less than $10^{-6}$ in each case.}
\label{table6}   
\end{table}

\section{Position resolution along the extruded scintillator}
\label{posres}
The experimental setup for this test is the same as shown in Fig.~\ref{fig:blackbox} except that $\sim$ 5\,m long WLS fibre is inserted in a ($60\,cm \times 5\,cm \times 2\,cm$) extruded scintillator. The WLS fibre is movable through the extruded scintillator to any position. The position where muon passes through the test scintillator is decided by a perpendicularly placed trigger paddle, P4 (width 5\,cm). P4 is placed in such a way that the position where muon passes through is nearly at the middle of the extruded scintillator. The extreme left of the test scintillator where the SiPM readout is 26.2\,cm from the coincidence area is assigned as a reference position, $x\:=\:0$. WLS fibre is then shifted towards the left from $x\:=\:0$\,cm to $x\:=\:250$\,cm with a gap of 50\,cm for each position.
\begin{figure} [htbp]  
\centering    
\includegraphics[height=8.5cm,width=14.75cm]{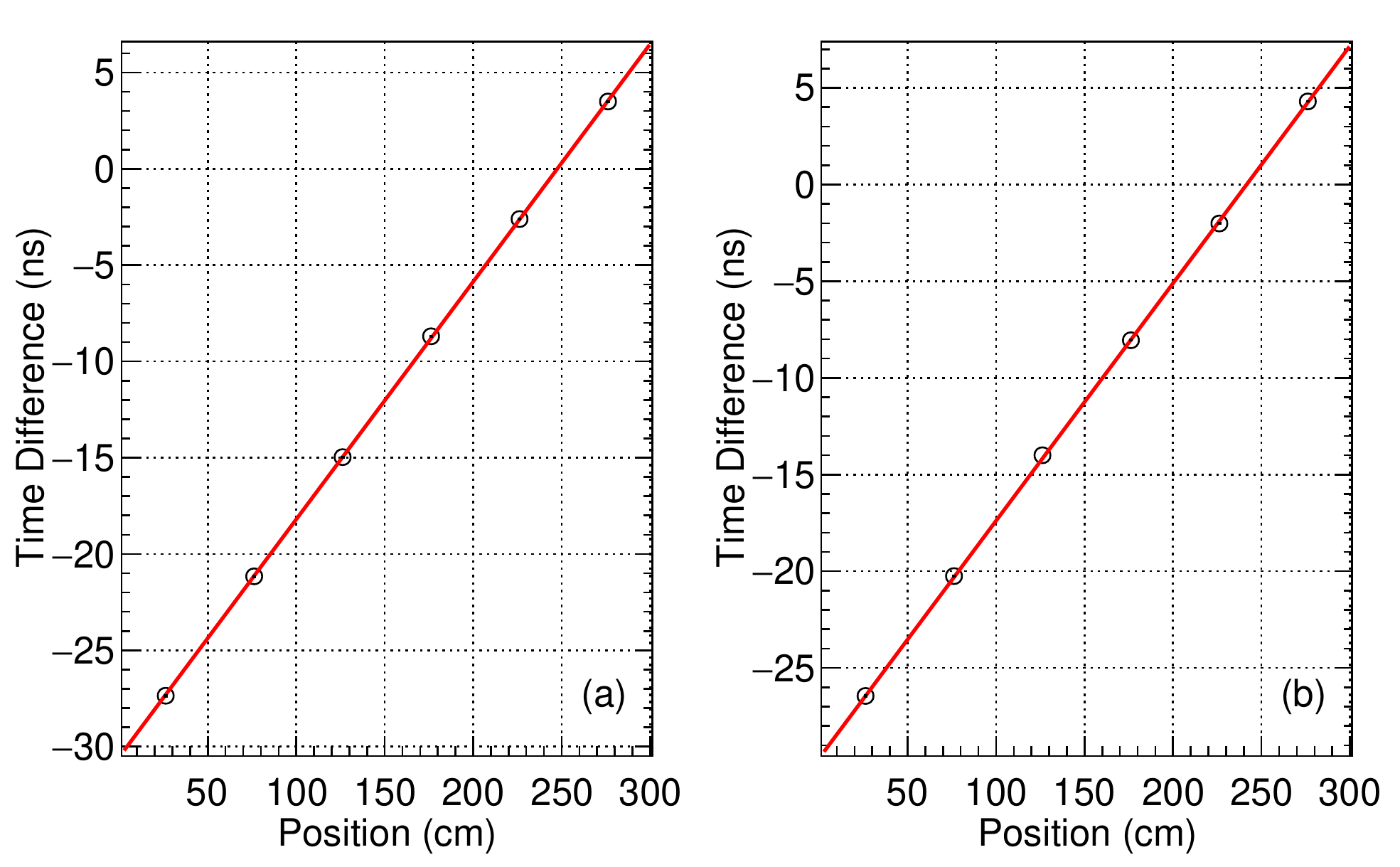}
\caption{The mean time difference as a function of muon position, (a) average of SiPMs on same sides, (b) average of SiPMs only on one side and only one SiPM on other side.}
\label{fig:posres}
\end{figure}
Overnight cosmic data are taken for a total of 6 positions, where the $V_{bias}$ to the SiPMs is maintained at 54\,V. The timing of each SiPM is calculated by measuring the time corresponding to the constant fractions (10\,\%, 20\,\%, 30\,\%, 40\,\%, 50\,\%) of the peak amplitude. The position resolution for this configuration can be measured by measuring the time difference of the SiPM signals of the opposite sides, as the timing of the SiPM is sensitive to the position of P4 and thus the location of muon inside the test scintillator~\cite{timeposres}. The time difference is calculated by taking the average time of two SiPMs on the same side. One of the SiPM responses is less as compared to other SiPMs due to misalignment with fibre, so to exclude the effect of this misaligned SiPM in the timing measurement, the timing difference was also calculated  by taking the average time of two SiPMs on one side, and considering only one SiPM on the other side. The mean time difference ($\Delta$t) at each position is plotted as a function of $x$, which is shown in Fig.~\ref{fig:posres} for the two cases mentioned above. Position resolution is calculated using the equation:
\begin{equation}
\label{eq:sigma}
\sigma_{x} = (\sigma_{\Delta t})_{avg} \times v
\end{equation}
where, $\sigma_{x}$ gives position resolution of the test scintillator, ($\sigma_{\Delta t}$)$_{avg}$ is the average value of standard deviation from time difference distribution of all positions and $v$ is the average speed of photon propagation inside the fibre. From the straight-line fit, the average speed of the photon inside the fibre measured from these two data sets is (16.287 $\pm$ 0.016)\,cm/ns. The average values of standard deviation from $\Delta\,t$ distributions are (2.019 $\pm$ 0.027)\,ns and (1.751 $\pm$ 0.019)\,ns for the two cases respectively. Consequently, the uncertainty on position measurement for the two cases,
(32.7 $\pm$ 0.5)\,cm and (28.5 $\pm$ 0.4)\,cm. This measurement is after subtracting the error due to muon position (due to width of P4), but that effect is negligible.

\section{Calibration of SiPM using a radioactive source}
\label{source}
The LED system will be used periodically for the calibration of the SiPMs. During the experimental runs it is very important to monitor the detector performance periodically and it can also be done using a radioactive source. This may not be the case in the CMV covering the mini-ICAL, but other systems can use a radioactive source to reduce the cost and complexity of electronics. So, for testing and characterizing the SiPM, the radioactive source, $^{22}Na$ is used with  $V_{ov}$ = 3\,V.

\begin{figure} [htbp]  
\centering
\includegraphics[height=7.25cm,width=11cm]{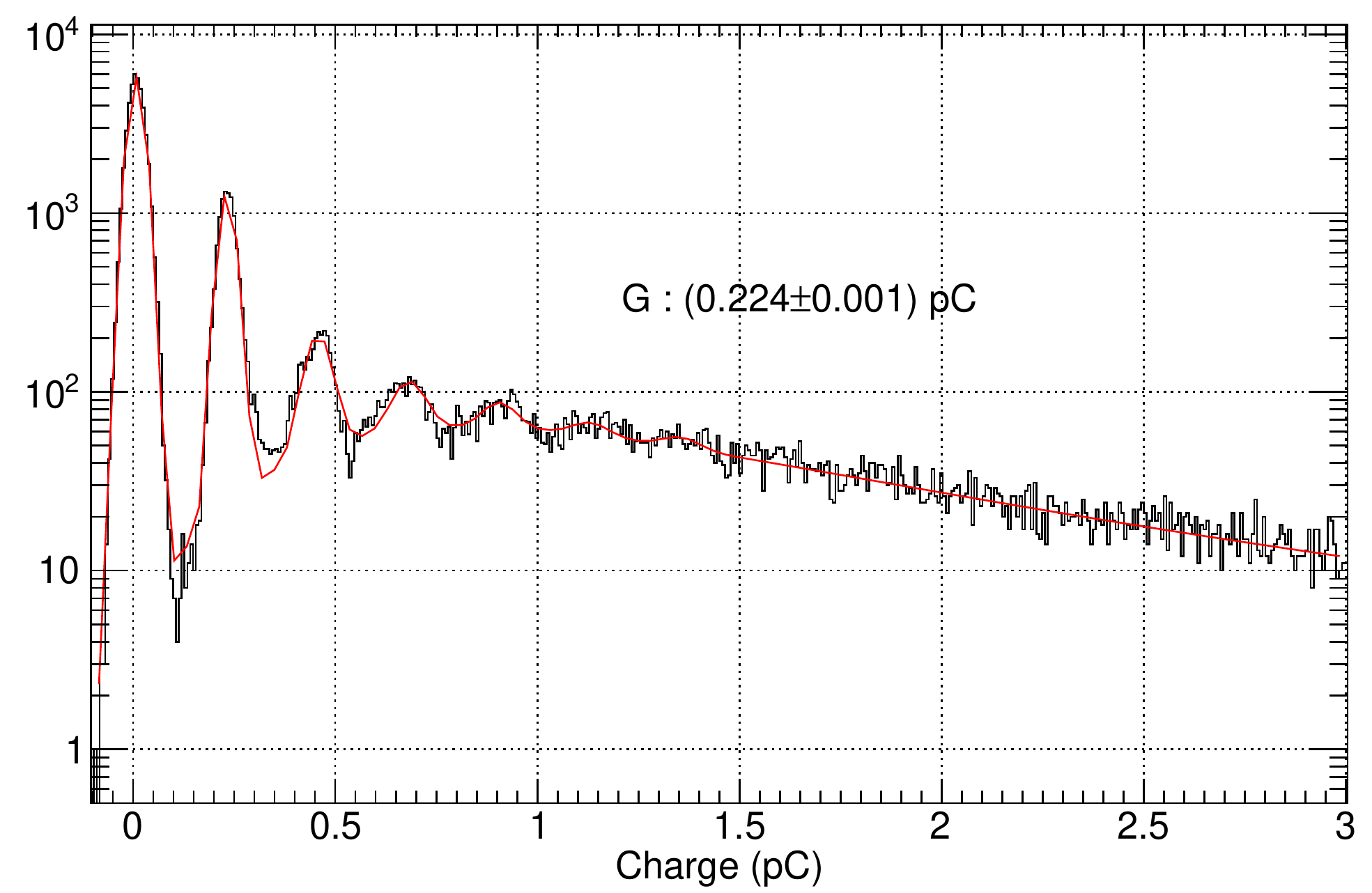}
\caption{Photoelectron peaks using a radio-active source in a SiPM.}
\label{fig:sipmpeaksource}
\end{figure}
The radioactive source $^{22}Na$ is kept on top of an extruded scintillator. The trigger is set on one of the SiPM channels with a threshold of slightly less than one photoelectron threshold. Whenever the trigger criterion is satisfied, data is accumulated through an oscilloscope for all SiPMs mounted on the extruded scintillator. The charge distribution with several photoelectron peaks is shown in Fig.~\ref{fig:sipmpeaksource}, where the data is fitted with Gaussian function for each peak and Landau function for the background.
The gain of the SiPM is evaluated by calculating the average gap between the consecutive peaks in the spectrum. Cosmic ray muon of energy about 2\,MeV produces $\sim$35 photoelectron in the SiPM. Thus from the Compton electron produced from 511\,keV photon, one can expect (0 - 6) photoelectron and which is what has been observed here. The long tail is due to the combination of 1.275\,MeV photons along with 511\,keV photons. 

\section{Calibration of SiPM using noise data}
\label{noisecalib}
\begin{figure} [htbp]  
\centering
\includegraphics[height=7.25cm,width=11.cm]{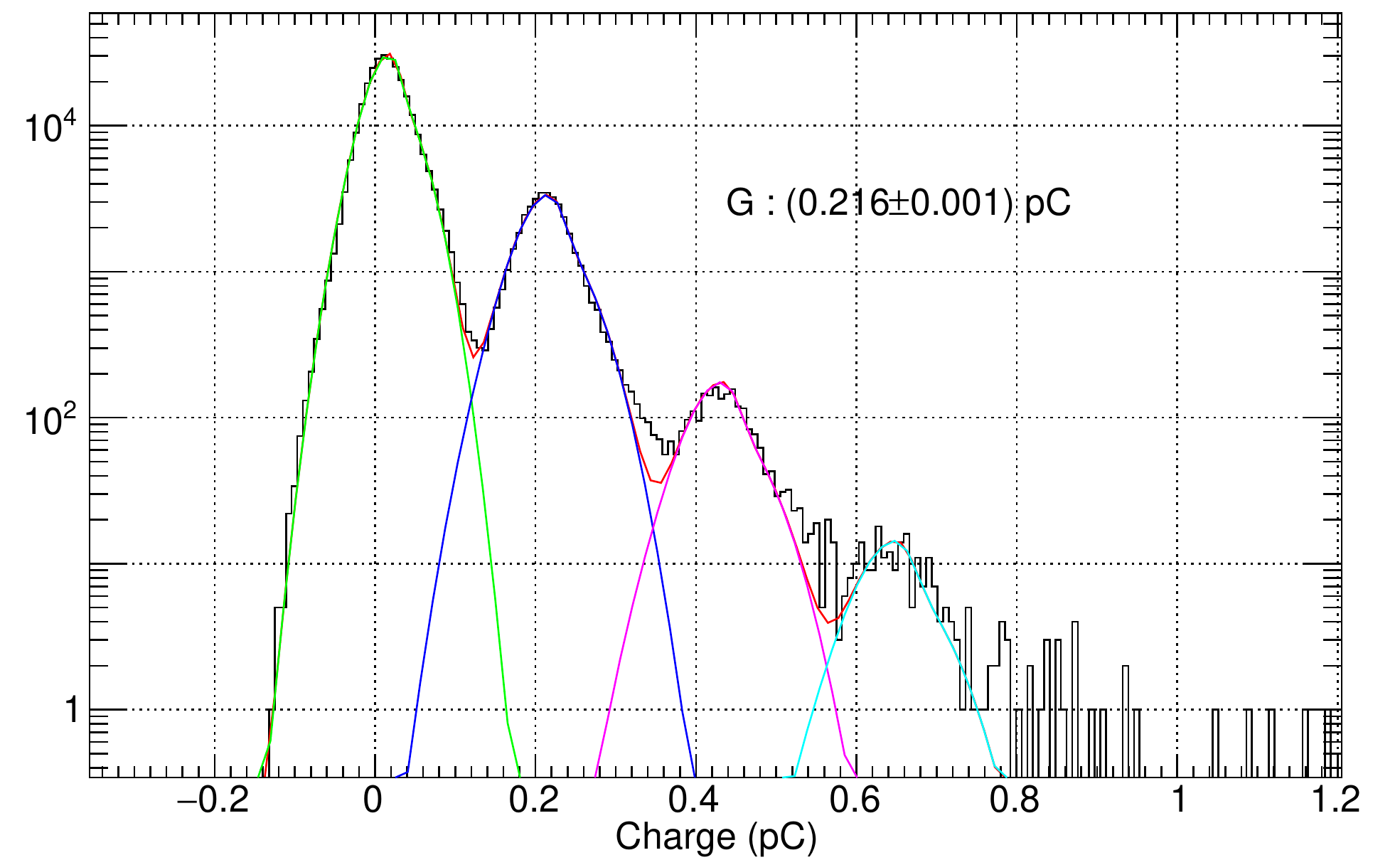}
\caption{Photoelectron peaks from noise trigger in an SiPM.}
\label{fig:sipmpeaknoise}
\end{figure}

Another alternative for calibrating SiPMs is by using noise data. Even though the accuracy is slightly poorer than that of the LED calibration, most of the experiments use this to calibrate SiPM. There are mainly two advantages in using this method neither is extra hardware needed nor there is any dead time. Events are triggered by a random signal from an external scintillator paddle. The integrated charge distribution with photoelectron peaks is shown in Fig.~\ref{fig:sipmpeaknoise}. This distribution is fitted with a sum of Gaussian functions to incorporate the pedestal and photoelectron peaks. Due to selection criteria, the pedestal is shifted slightly on the positive side and the gain of the SiPM is evaluated by calculating the average gap between the consecutive photoelectron peaks in the spectra. The single photoelectron gain estimation from both LED source and $^{22}Na$ source data agree well with this data. 

\section{Correlated noise in SiPM}
\label{correlated}
Various parameters deteriorate the performance of the SiPMs, e.g., primary Dark Count Rate (DCR) or primary noise rate as well as correlated noise like crosstalk and afterpulse. One of the common methods to quantify these parameters is to analyze the output waveforms from SiPM in a dark environment at a controlled temperature~\cite{sipmpaper4,sipmpaper5}, which is also studied here. To measure the above parameters, the waveforms from SiPM are amplified with a voltage amplifier of gain 25. The amplified waveforms are stored at $V_{ov}\:=\:(1.5$ - $6)\,V$ with a gap of 0.5\,V using a self-triggering threshold of less than one photoelectron peak.

\begin{figure} [htbp] 
\centering
\includegraphics[height=8.cm,width=12cm]{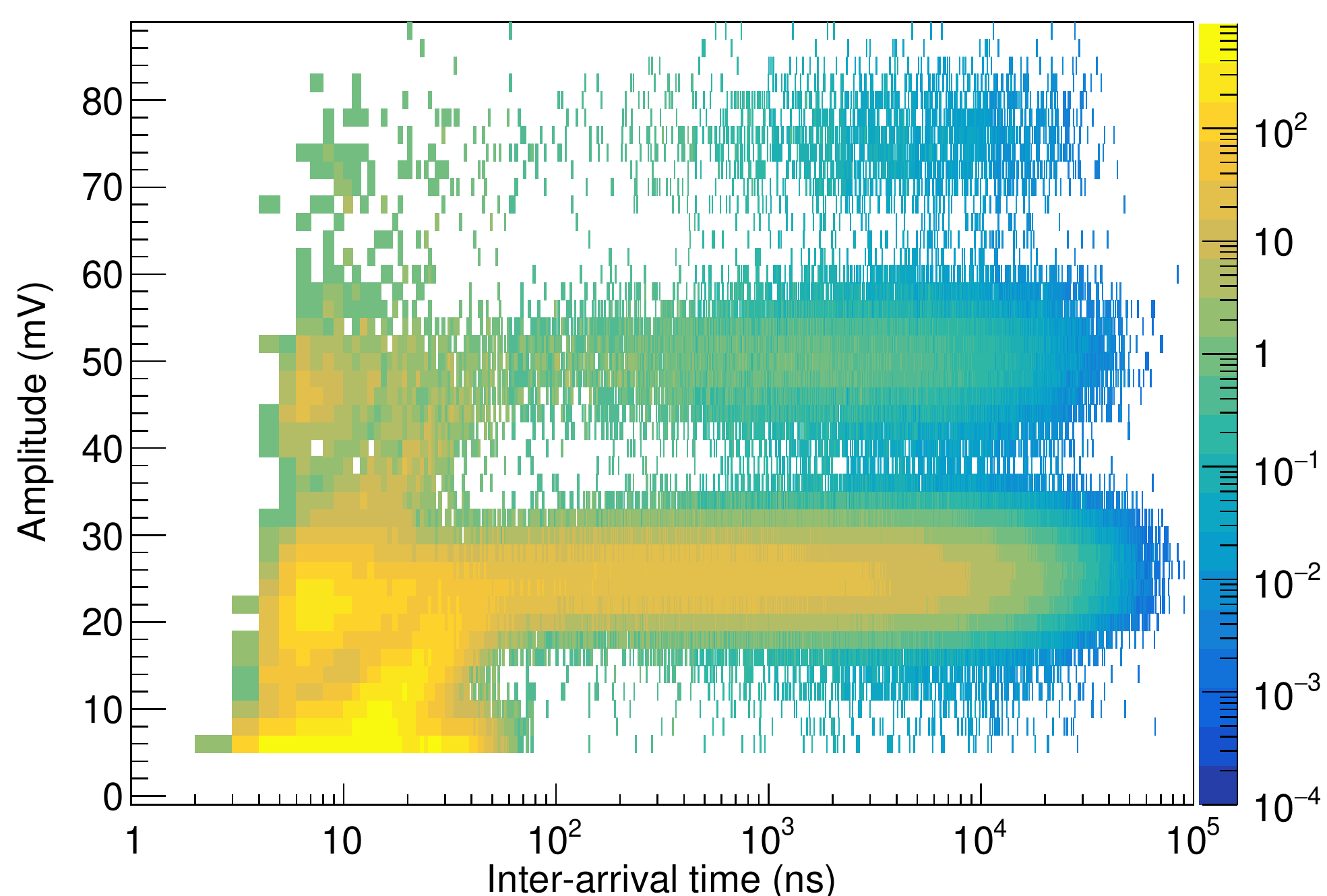}
\caption{Amplitude of second pulse as a function of time difference ($\Delta$t) of SiPM signal at $V_{ov}\:=\:3\,V$. }
\label{fig:crosstalk2D}
\end{figure}  

\begin{figure}[htbp]
\captionsetup[subfigure]{labelformat=empty}
\centering
\hspace*{-0.5cm}
\begin{subfigure}{0.42\textwidth}
\centering
\includegraphics[height=5.cm,width=6.cm]{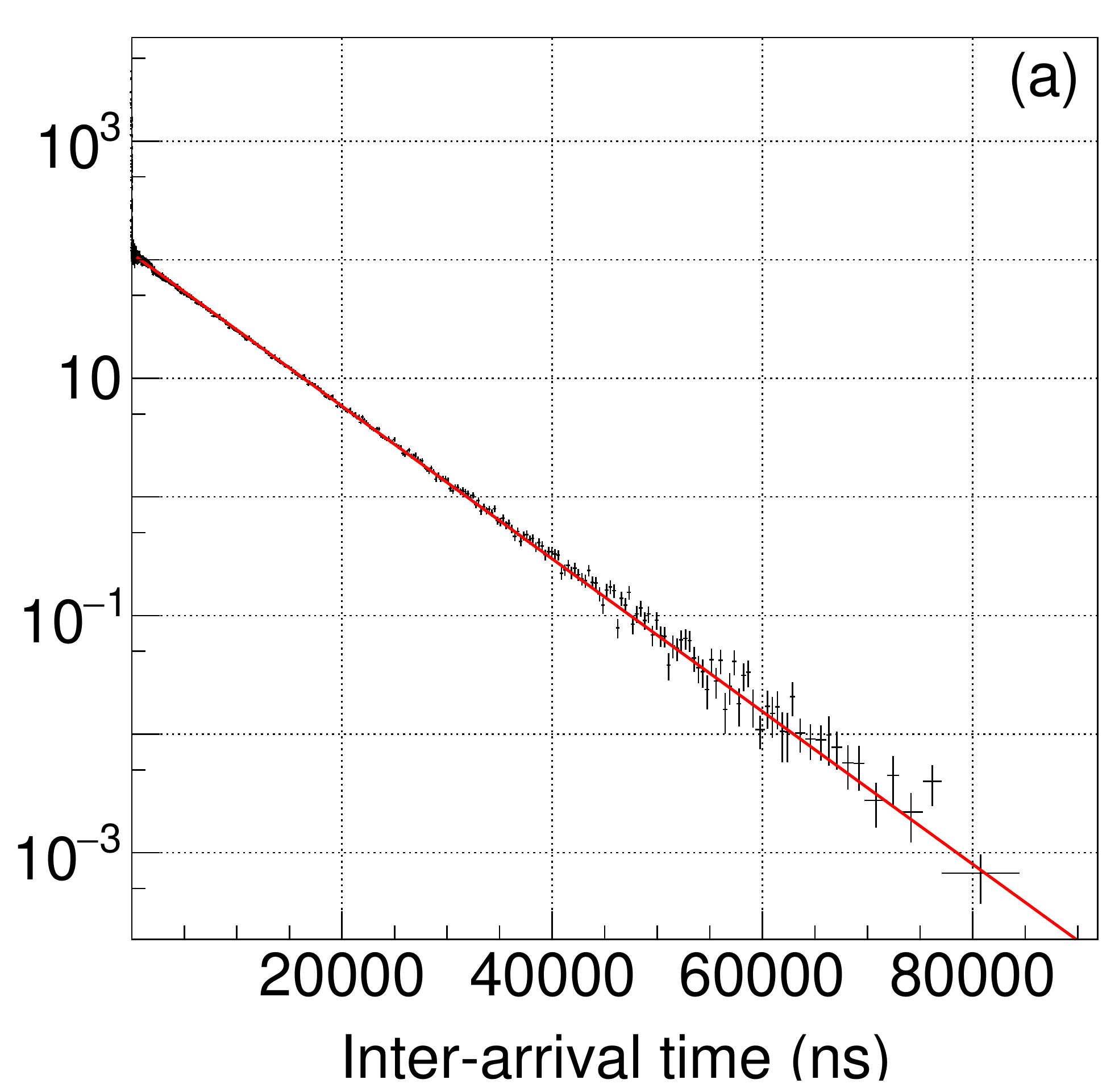}
\caption{}
\label{fig:semilog}
\end{subfigure}
\hspace{0.5cm}
\begin{subfigure}{0.42\textwidth}
\centering
\includegraphics[height=5.cm,width=6.cm]{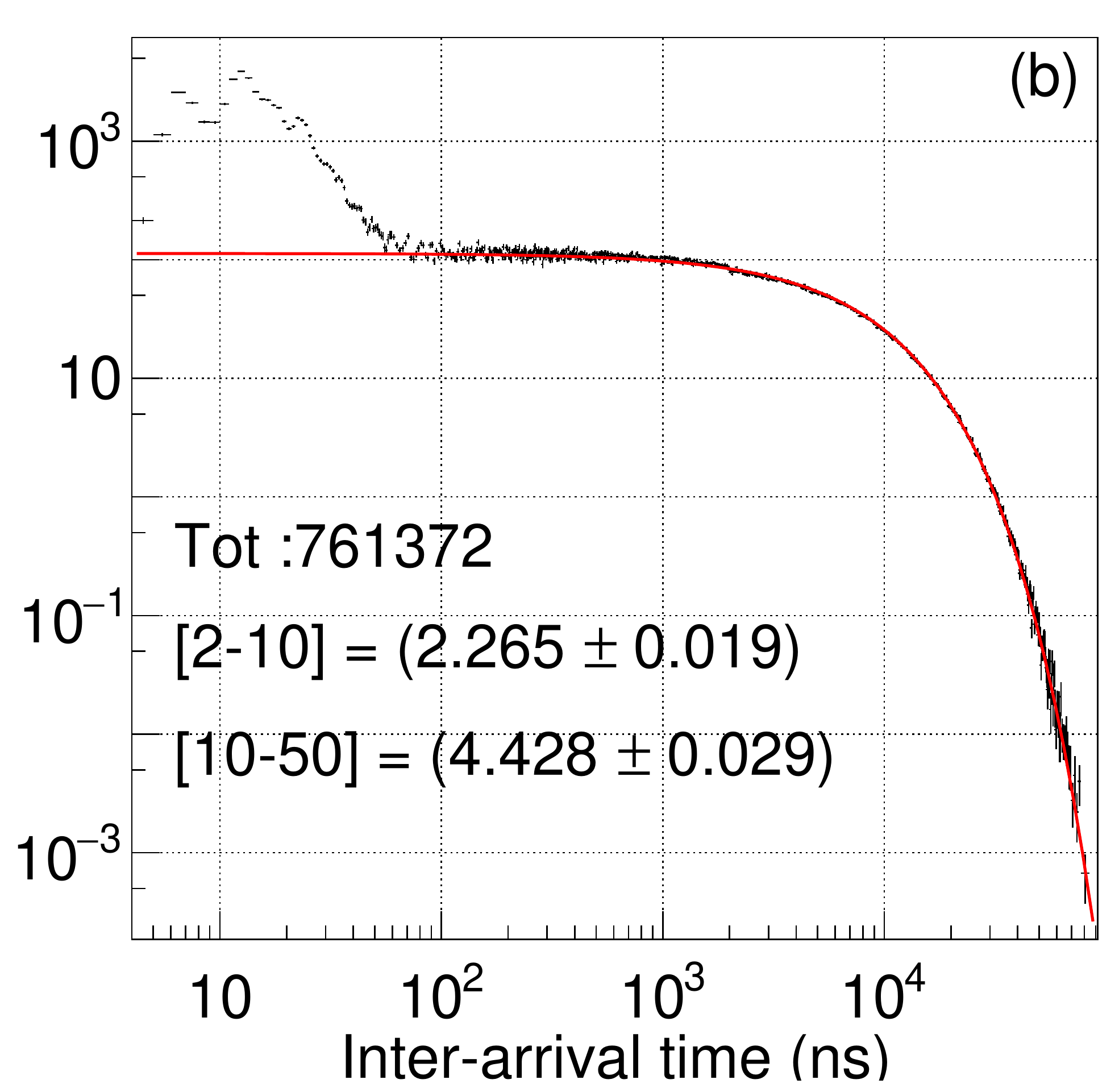}
\caption{}
\label{fig:log}
\end{subfigure}
\caption{Time difference ($\Delta$t) of primary noise rate as well as correlated noise rates with the fit of exponential function which represent the random chance coincidence in (a) semilog scale and (b) log-log scale. }
\label{fig:crosstalk1D}
\end{figure}
The rate of the direct crosstalk increases the amplitude of the pulse which can be distinguished from primary noise counts. On the other hand, afterpulse and delayed crosstalk counts will be merged with the primary noise counts. The method used here requires a time difference between the consecutive pulses and peak amplitude corresponding to the second pulse. There is a separation of afterpulse and delayed crosstalk counts from primary noise counts which are clearly visible in the amplitude versus time difference plot as shown in Fig.~\ref{fig:crosstalk2D}. To calculate primary DCR, time difference distribution is fitted exponentially in a range where the time difference is more i.e., no afterpulse and delayed crosstalk events are present as shown in Fig.~\ref{fig:semilog}. Then by extrapolating the exponential fit to the lower side of the distribution, the difference between measured and extrapolated values are used to evaluate the afterpulse ([10 - 50]\,ns time window) and crosstalk ([2 - 10]\,ns time window) rate as shown in Fig.~\ref{fig:log}.% , which is similar to other publications. % GMA Give the reference for my slides, same NIM paper etc

\begin{figure}[htbp]
\captionsetup[subfigure]{labelformat=empty}
\centering
\hspace*{-0.5cm}
\begin{subfigure}{0.42\textwidth}
\centering
\includegraphics[height=5.cm,width=5.cm]{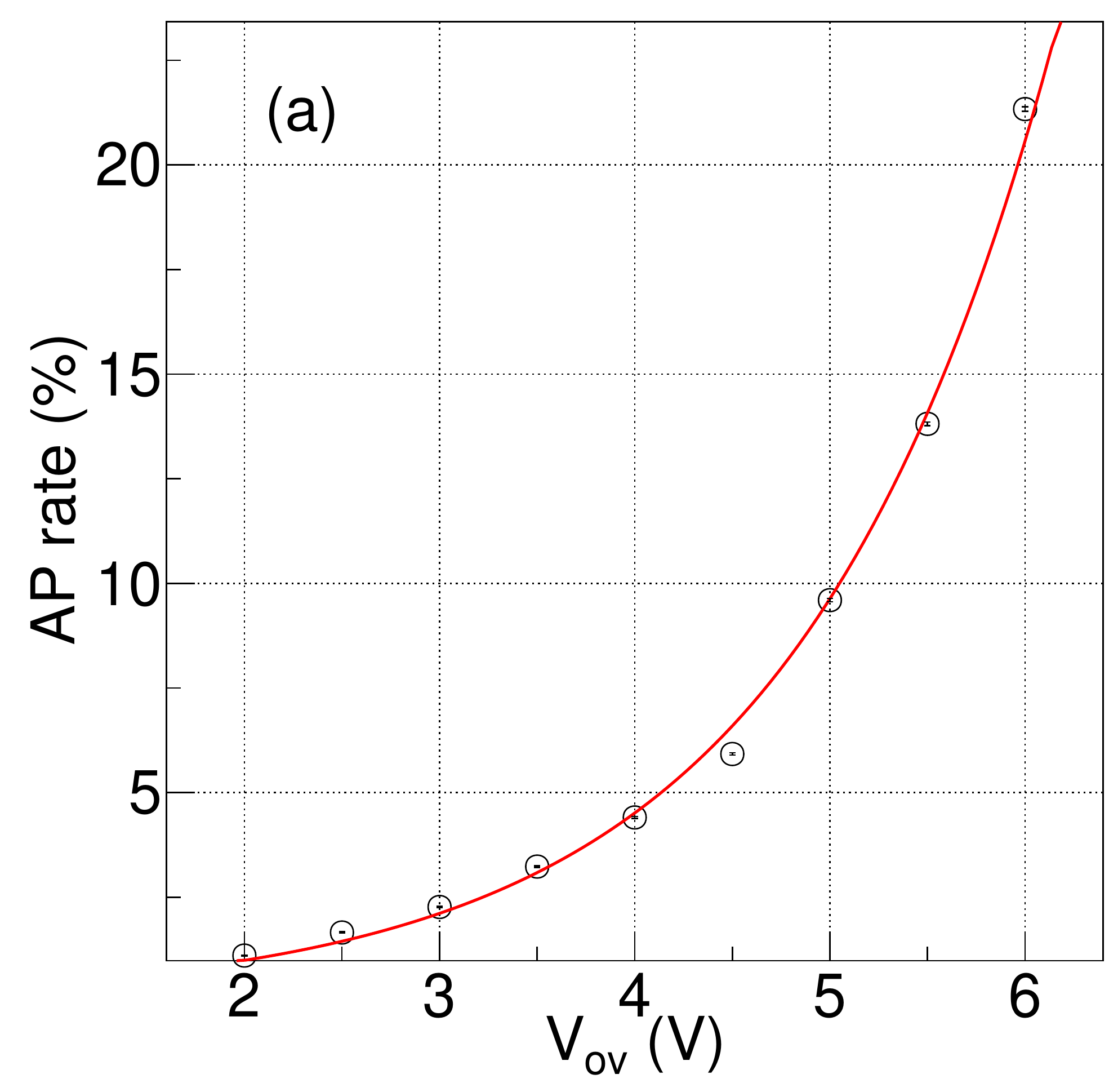}
\caption{}
\label{fig:AP}
\end{subfigure}
\hspace{0.5cm}
\begin{subfigure}{0.42\textwidth}
\centering
\includegraphics[height=5.cm,width=5.cm]{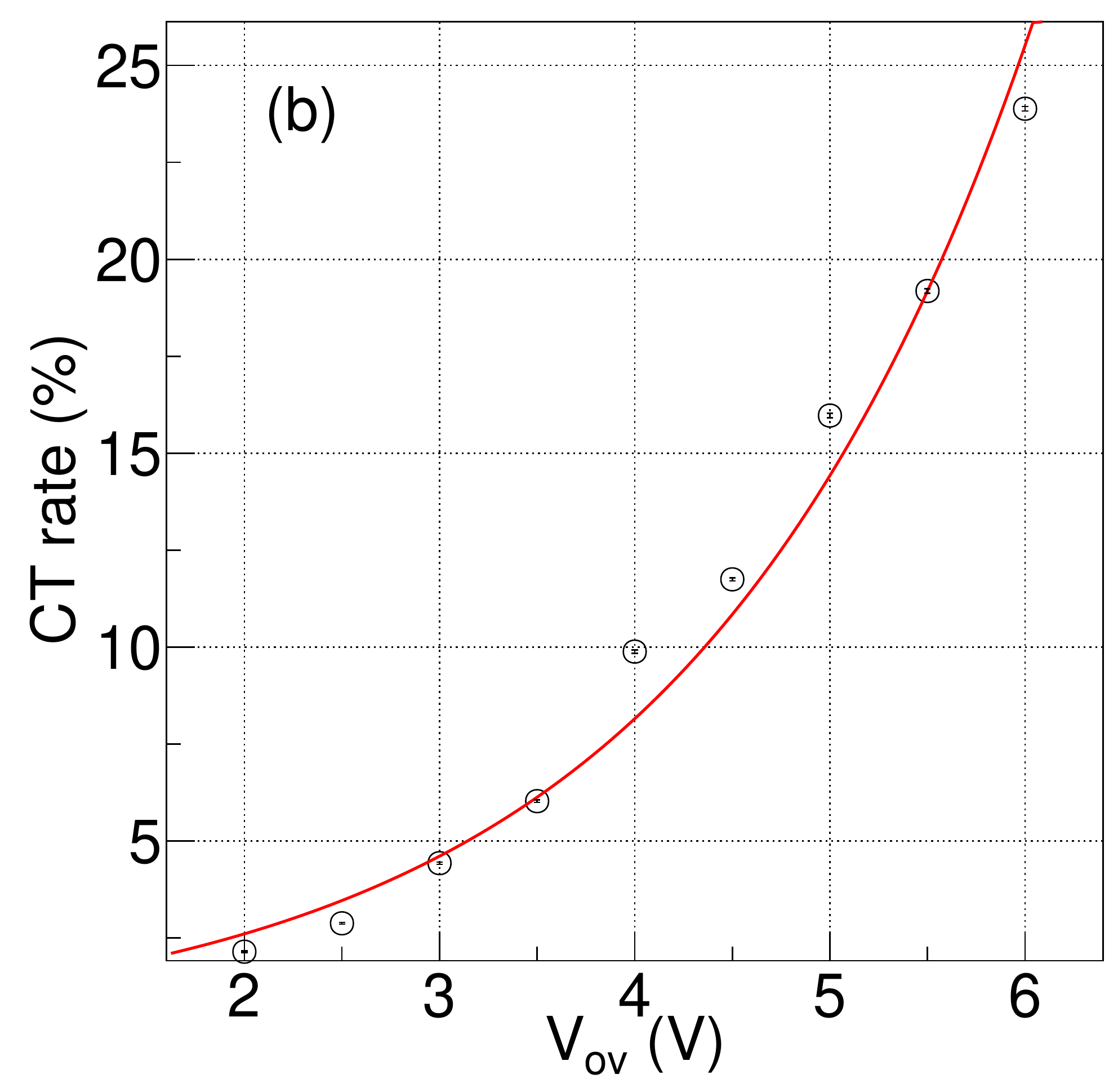}
\caption{}
\label{fig:CT}
\end{subfigure}
\caption{Correlated noise rate in a SiPM, (a) Afterpulse and (b) Crosstalk as a function of $V_{ov}$. }
\label{fig:AP_CT}
\end{figure}

The afterpulse and crosstalk rates for different $V_{ov}$ are shown in Fig.~\ref{fig:AP_CT}, which are consistent with the specification of these SiPMs.

\section{Recovery time of SiPM}
\label{recovery}
Recovery time ($\tau$) is the time taken by an SiPM to regain the original state after an avalanche. The recovery time of SiPM is calculated using the waveform analysis method. Amplified waveforms (with gain 25) from SiPMs are collected using a double pulse trigger on an oscilloscope. The trigger is generated when there are two pulses within 70\,ns.\\ 
\begin{figure}[htbp]
\captionsetup[subfigure]{labelformat=empty}
\centering
\begin{subfigure}{0.325\textwidth}
\includegraphics[height=4.90cm,width=4.90cm]{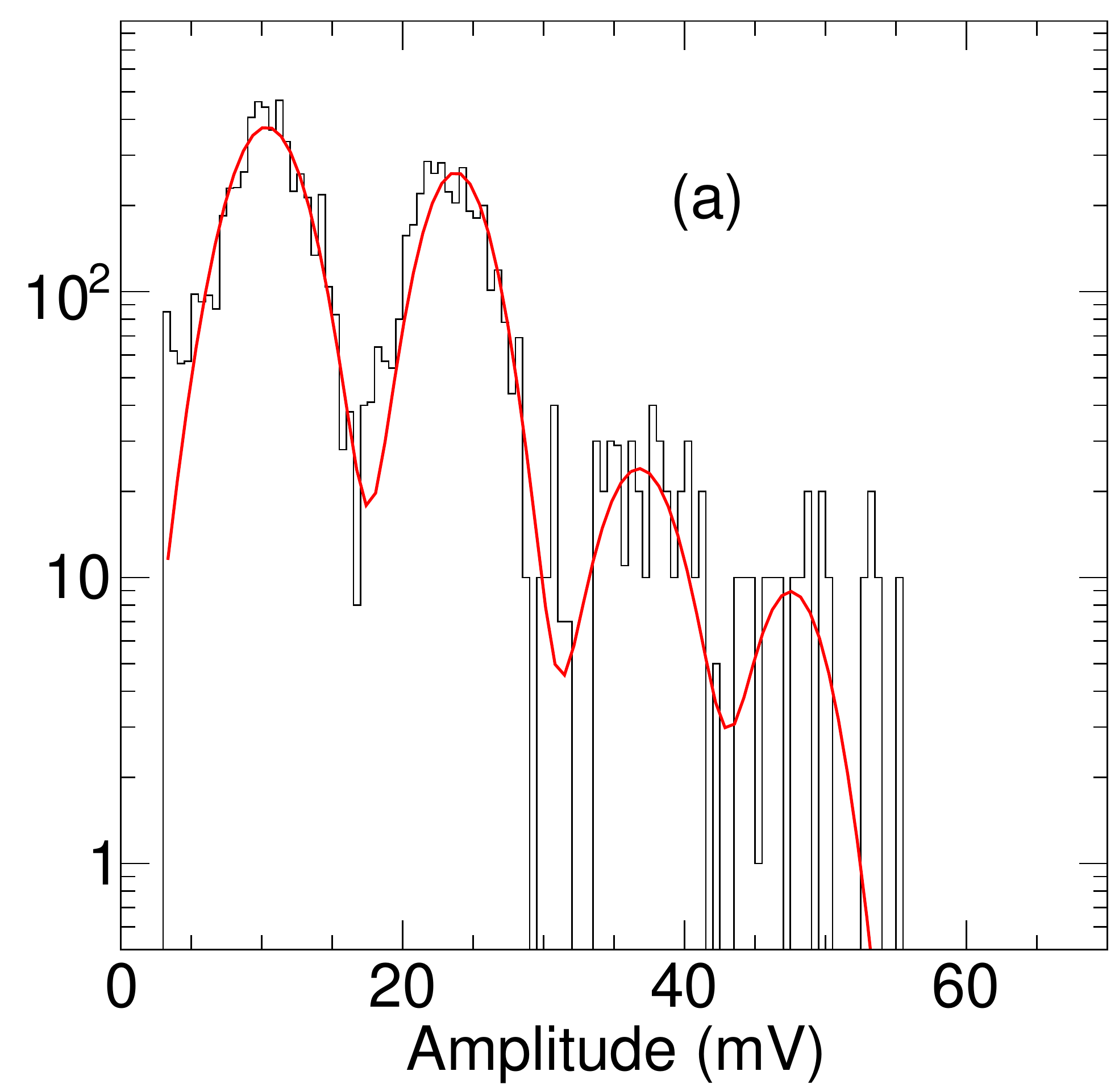}
\caption{}
\label{fig:4peaks}
\end{subfigure}
\begin{subfigure}{0.325\textwidth}
\includegraphics[height=4.90cm,width=4.90cm]{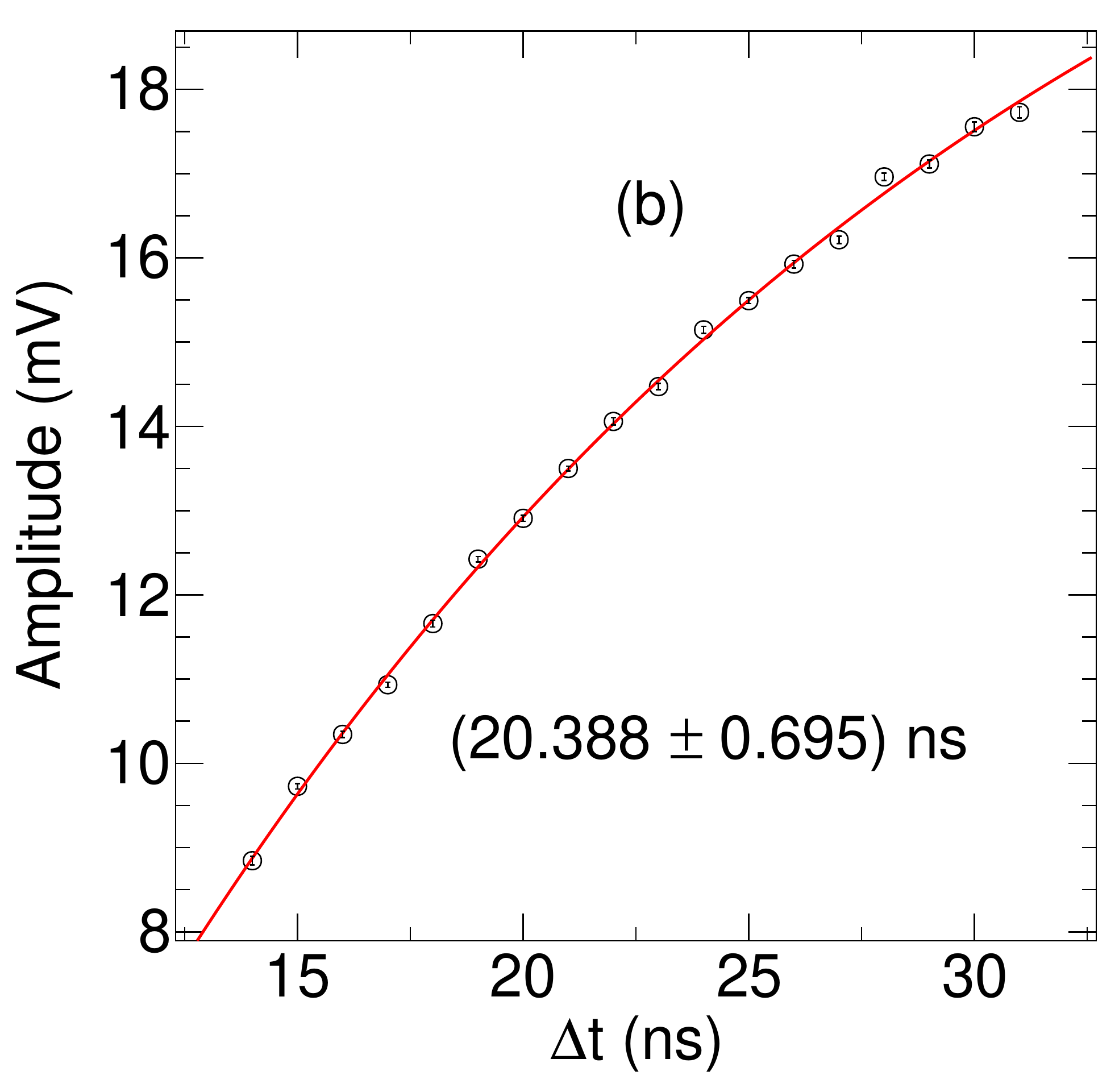}
\caption{}
\label{fig:discharge}
\end{subfigure}
\begin{subfigure}{0.325\textwidth}
\includegraphics[height=4.90cm,width=4.90cm]{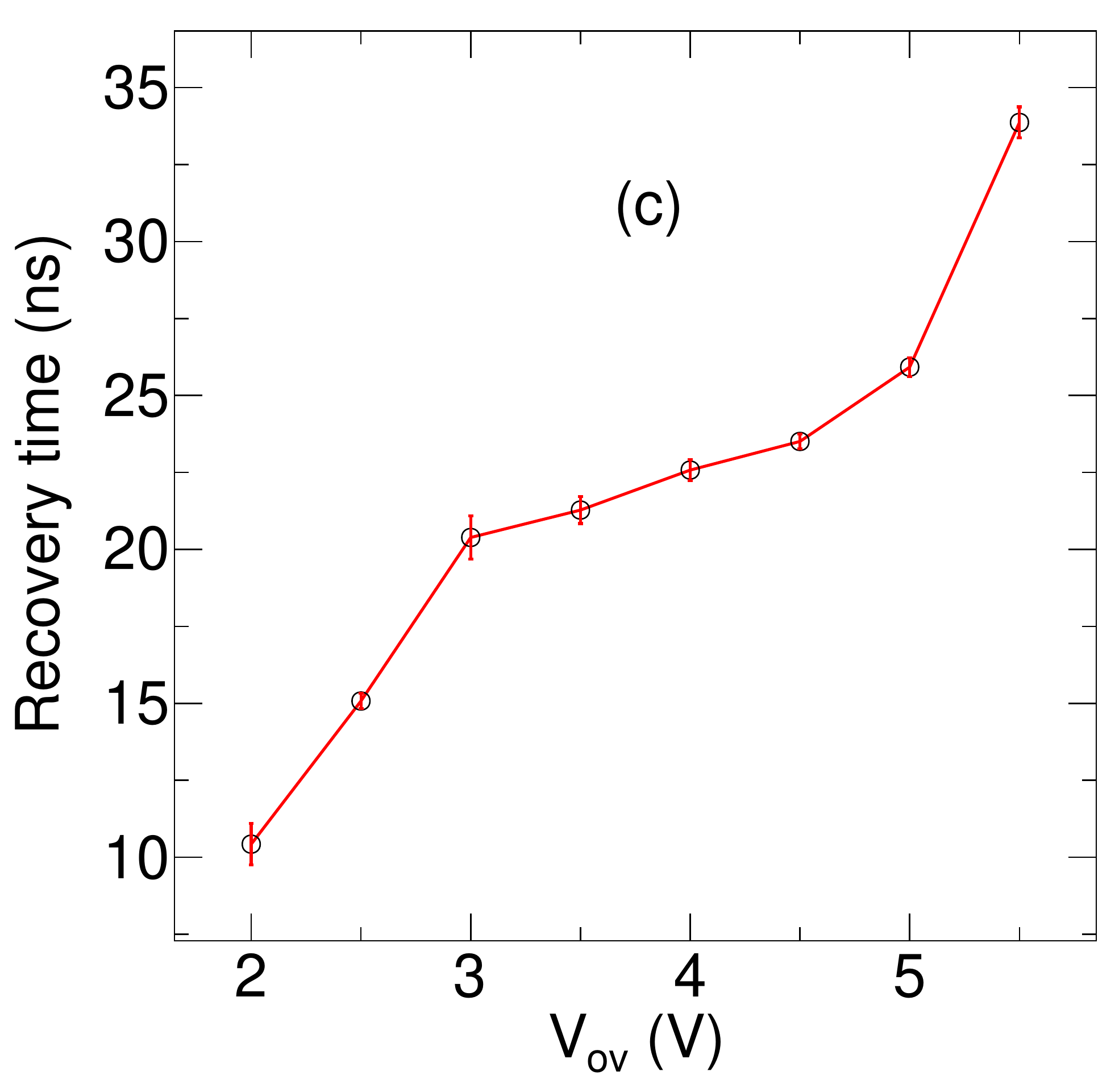}
\caption{}
\label{fig:tau}
\end{subfigure}
\caption{(a) Amplitude of the second pulse for $\Delta t$=16-17\,ns at $V_{ov}$=3\,V, (b) Mean amplitude of after pulse as a function of $\Delta t$ at $V_{ov}$=3\,V and (c) Variation in recovery time for different $V_{ov}$.}
\label{fig:recovery}
\end{figure}

As mentioned in the Section~\ref{correlated}, amplitude versus time difference distribution is plotted similar to what is shown in Fig.~\ref{fig:crosstalk2D} except the x-axis range is (0-100)\,ns. The pulse height of the second peak for $\Delta t\:=\:(16$ - $17)\,ns$ is shown in Fig.~\ref{fig:4peaks}, where the four peaks are due to the first afterpulse, single photoelectron peak, second afterpulse, and the second photoelectron peak respectively. This distribution is fitted with four Gaussian functions, where the mean and width of photoelectron peaks are fixed from data beyond 100\,ns of Fig.~\ref{fig:crosstalk2D}. For an afterpulse signal, widths are floated but the distance of a peak position from the next photoelectron peaks is kept the same. This was repeated for different time windows. Fig.~\ref{fig:discharge} shows the mean amplitude of afterpulse as a function of the time difference w.r.t first pulse which is fitted with the function,

\begin{equation}
\label{eq:dischargeeq}
f(\Delta t) = A \times\Bigg[1 - exp\bigg(-\frac{(\Delta t - t_0)}{\tau}\bigg)\Bigg]
\end{equation}
where $\tau$ is the recovery time of the SiPM. All these measurements are done at $V_{ov}\:=\:(2$ - $6)\,V$ with a gap of 0.5\,V. Fig.~\ref{fig:tau} shows the variation in recovery time as a function of $V_{ov}$, and as expected for a larger pulse, it takes a longer time to recover~\cite{sipmpaper5}. This behaviour and the magnitudes are also observed in other studies~\cite{recovery}.  

\section{Conclusion}
\label{conclusion}
Main goals of this study was to characterize the SiPM, make a testbench to test the quality of all SiPMs used for the muon veto detector, obtain the optimum $V_{ov}$ and the threshold values of integrated charge in each SiPM. Even a more important goal was to arrive at the muon veto criteria for this experiment. Starting from the most common method i.e., LED calibration, the gain of the SiPM is measured as a function of $V_{ov}$. Apart from the LED method, two alternative methods of calibration have been established, i.e., using radioactive source data and noise data. The gains derived from these three procedures were seem to be consistent with each other. \\
To decide on the optimum $V_{ov}$ and $q_{th}$, cosmic muon efficiency and noise rate for a single extruded scintillator as well as for the layers (i.e., the whole detector system) are measured using a 1\,cm thick extruded scintillator. For the CMV detector, both 1\,cm as well as 2\,cm thick extruded scintillators will be used. The efficiency requirements that are satisfied by 1\,cm thick extruded scintillators will also be satisfied by 2\,cm thick extruded scintillators. Based on the above results, the optimum operating voltage, $V_{ov}$, is found to be 2 to 2.5\,V and the threshold value of the integrated charge is 0.4 pC. Finally, the veto efficiency will be measured by taking coincidence of any two out of four SiPMs from each scintillator and any two out of four scintillator layers.\\
Correlated noise of SiPM is observed to be (5 - 6)$\%$ (at $V_{ov}$=3\,V) of the total noise which is in agreement with it's specifications. The recovery time of SiPM is found to vary between (10 - 35)\,ns at different values of $V_{ov}$.

\section{Acknowledgments}
We sincerely thank Piyush Verma, S.R. Joshi, Darshana Gonji, Santosh Chavan, Vishal Asgolkar and Pathaleswar Esha for their support and help during the work. We would also like to thank all other members of the INO collaboration for their valuable inputs. A special thanks to Eric A. Fernandez from Fermilab for sharing his knowledge about extruded scintillator production.

\end{document}